\documentclass[conference]{IEEEtran}
\IEEEoverridecommandlockouts
\usepackage{cite}
\usepackage[size=smaller]{caption}
\usepackage{amsmath,amssymb,amsfonts, bbm}
\usepackage{subcaption}
\usepackage{stfloats} 
\usepackage{booktabs} 
\usepackage{graphicx}
\usepackage{textcomp}
\usepackage{xcolor}
\usepackage{array}
\usepackage{multirow}

\usepackage[ruled,vlined]{algorithm2e}
\usepackage{multicol}
\usepackage{upgreek}

\def\BibTeX{{\rm B\kern-.05em{\sc i\kern-.025em b}\kern-.08em
    T\kern-.1667em\lower.7ex\hbox{E}\kern-.125emX}}
\begin{document}

\title{
Driving up Inference Energy on SNNs:\\Per-Sample and Universal Sponge Attacks
}



\author{
    \IEEEauthorblockN{
        Spyridon Raptis and Haralampos-G. Stratigopoulos}
    \IEEEauthorblockA{\small{
        Sorbonne Université, CNRS, LIP6, Paris, France}}}

\maketitle

\begin{abstract}
Spiking Neural Networks (SNNs) communicate through sparse binary spike events rather than 
dense activations, enabling energy-efficient inference on neuromorphic hardware and motivating 
their use in always-on, battery-powered edge systems. We show that this same efficiency 
advantage creates a distinct security risk: \emph{sponge attacks} can increase inference-time 
spike activity and synaptic workload, inflating energy consumption while remaining difficult to 
detect through correctness-based monitoring alone.
Prior input-space efficiency attacks on SNNs have focused on per-sample optimization, primarily
in rate-coded settings. We extend this threat to native event-based binary inputs and study two
attack models. First, we develop a \emph{per-sample sponge attack} that crafts a custom
adversarial spike train for each input via gradient-based optimization. This attack increases per-inference SynOps by
$1.5$--$2.6\times$, on three SNN models for NMNIST, SHD, and IBM DVS Gesture datasets, while preserving the predicted class on at least $98\%$ of
evaluated samples. Second, to the best of our knowledge, we introduce the first
\emph{universal sponge attack} for native
event-based SNN inputs: a fixed binary perturbation computed offline and applied via XOR to all
subsequent inputs. Although weaker, it still inflates SynOps by $1.09$--$1.24\times$ across all
three datasets and represents a more realistic deployment threat because it requires no per-input
optimization.
Mapping SynOp inflation to estimated Loihi-1 energy yields per-inference overheads from
$14\,\upmu\mathrm{J}$ to $13.24\,\mathrm{mJ}$. These results show that native event-based SNNs
are vulnerable to practical input-space efficiency attacks, and that reusable universal
perturbations can accumulate into meaningful battery drain in continuously deployed edge systems.
\end{abstract}


\section{Introduction} \label{sec:introduction}
The energy a Spiking Neural Network (SNN) consumes at inference time
is, to a first approximation, proportional to the number of spikes
that propagate through the network. On event-driven neuromorphic
hardware---Intel's Loihi~\cite{Loihi18}, IBM's
TrueNorth~\cite{ASCA-IAMINDNTBBKMRJ15} and similar
chips---no spike means no synaptic operation and therefore (almost)
no energy. This is precisely what makes SNNs attractive for
always-on, battery-powered edge sensing: event-based vision (e.g., AR glasses, drones, security cameras), environmental sound classification, wearable health monitoring (e.g., ECG, EEG, EMG), wake-word and keyword spotting, robotics (e.g., obstacle detection \& collision avoidance, visual navigation), predictive maintenance and anomaly detection, etc.
The same proportionality, however, exposes a security property that
has no clean analogue in dense neural networks. An adversary who can
perturb the
input to a deployed SNN can drive its spike activity arbitrarily
close to its saturation point, draining the device's battery, while
the network's prediction remains correct.

We refer to such inputs as \emph{sponge inputs} after Shumailov et~al.
~\cite{shumailov2021sponge}, who introduced the term in the Artificial Neural Network
(ANN) setting. A sponge attack on an SNN succeeds when (i) the
network's predicted class for the adversarial input matches the
predicted class for the clean input, and (ii) the number of
spike-driven synaptic operations consumed per inference is
substantially higher. The operational consequence is a device whose
user observes the model \emph{behaving normally}---correct
predictions, similar latency---but whose battery drains faster than
expected. 

While sponge attacks on conventional ANNs have been investigated under several
threat models, including run-time inputs~\cite{shumailov2021sponge}
and training-time poisoning~\cite{cina2022sponge}, the SNN-specific
landscape is sparse. To our knowledge, the only published
input-space sponge attack against SNNs is
\textit{SpikeAttack}~\cite{krithivasan2022spikeattack}, which evaluates on
two non-spiking image-classification benchmarks (CIFAR-10 and
ImageNet) by rate-coding pixel intensities into Poisson spike
trains and optimizing a continuous additive perturbation under an
$L_p$-norm budget on the (continuous) image. \textit{SpikeAttack} does not
evaluate on any native event-based dataset, and it operates exclusively per-sample.

The setting that makes SNNs operationally valuable is the one in
which the input is itself a stream of binary events produced by an
event camera or a silicon cochlea, represented as a binary
spatio-temporal tensor.
This regime differs from rate-coded image classification in three
ways. The input alphabet is binary, so an $L_p$-ball constraint
collapses to an $L_0$ flip count. Clean inputs are extremely sparse,
and time matters explicitly: the same number of events distributed differently along
$T$ produces a different cost profile through the network. Whether
input-space sponge attacks transfer to this regime, and at what
cost, is open.

\paragraph*{Two threat models, one shared cost target}

We answer this question by implementing and evaluating two
qualitatively different attack regimes that share the same harm
model (inflate per-inference cost while preserving the predicted
class):
\begin{itemize}
\item \textbf{Per-sample sponge attack:}
For each clean input we run a gradient
optimization that produces a custom binary adversarial spike train.
This is the analogue of \textit{SpikeAttack} adapted to the binary event
regime. It produces large inflation ratios but its per-input
optimization cost is prohibitive for streaming deployment: an
adversary would need to interpose the sponge generation compute between the
sensor and the SNN for every inference.
\item \textbf{Universal sponge attack:}
We construct a single fixed binary perturbation
$\mathbf{U}$, offline from a small
training-set subset (one-shot, $\sim\!1$ minute), then apply it to
every subsequent input by elementwise XOR. There is no per-input
optimization on the deployed device; injecting $\mathbf{U}$ at the
sensor preprocessing stage is enough. We show this is the
\emph{realistic operational threat}: a one-time injection
that requires neither attacker compute on the sensor pipeline nor
attacker access during inference.
\end{itemize}

Hence we make the following contributions:
\begin{enumerate}
\item We formulate and implement the input-space sponge attack
problem in the native binary event-based regime, the regime in
which SNNs are actually operationally valuable.
\item We show that the per-sample attack works on three
SNN victims (NMNIST \cite{OJCT15}, SHD \cite{heidelberg}, IBM DVS Gesture \cite{ATBM17}), that derive from both
vision and auditory domains, with spike-count inflation
of $1.8$ to $2.46\times$ while preserving the predicted class in $\geq 98\%$ of the cases.
\item We propose a universal sponge attack that inflates 
Synaptic Operations (SynOps) by
$1.09$ to $1.24\times$ across the same three datasets while
requiring only one-time construction and constant-time XOR at
inference. To the best
of our knowledge this is the first universal sponge attack
reported for SNNs in any regime, and the first universal sponge
attack to target the \emph{forward pass} of any neural network
class: existing universal-sponge attempts on dense
ANNs~\cite{shapira2023phantom,hong2021earlyexit} target
architectural quirks (non-maximum suppression in object detectors;
multi-exit branch logic in early-exit networks) rather than the
dense forward pass itself, which our analysis suggests is hard to
attack universally because the dense MAC count is essentially
input-independent (Section~\ref{sec:sota}).
\item We translate measured SynOp counts to estimated per-inference
Loihi-1 energy~\cite{Loihi18} and report absolute
$\Delta E$ figures (Section~\ref{sec:results-energy}). The absolute
energy delta scales with input dimensionality even when the
inflation ratio is modest, making the IBM DVS Gesture victim
($\sim 9\,\mathrm{mJ}$ clean baseline on Loihi 1) the most
operationally costly target in absolute terms.
\end{enumerate}

The rest of the paper is organized as follows.
Section~\ref{sec:sota} reviews related work on sponge attacks
against ANNs and SNNs.
Section~\ref{sec:threat} defines the threat model and the two
deployment-time attacker capabilities we consider.
Section~\ref{sec:notation} introduces the SNN notation and the
activity, cost, preservation, and stealth metrics used throughout.
Section~\ref{sec:method} presents the per-sample and universal
sponge attacks.
Section~\ref{sec:setup} describes the three victim SNNs, datasets,
and the attack configurations.
Section~\ref{sec:results} reports the per-sample and universal
sponge results, the translation to per-inference Loihi-1 energy,
and the attack runtime.
Section~\ref{sec:conclusion} concludes the paper.
\section{State of the Art}\label{sec:sota}

We organize the related work into four strands: sponge attacks against dense ANNs, 
universal sponge attacks against dense ANNs, input-space sponge attacks against SNNs, 
and parameter-space attacks against SNN energy.

\subsection{Sponge attacks on dense ANNs}\label{sec:sota-ann}
Shumailov et al.~\cite{shumailov2021sponge} introduced
\emph{sponge examples}: inputs deliberately crafted to maximize the
cost of a forward pass through a deployed ANN while preserving its
predicted class. Their formulation exploits two mechanisms:
input-dependent cost structures in Natural Language Processing (NLP) models (sequence length,
embedding density), and reduced activation sparsity in vision
models on accelerators that exploit zero-skipping. They report
energy and latency inflation factors of roughly $30\times$ on
white-box NLP models, and a black-box case study against a
commercial translation service in which a single sponge input
drove inference latency from $\sim\!1\,\mathrm{ms}$ to several
seconds. The threat model is run-time and white-box (or transfer-
based black-box): the adversary perturbs the model's input but
not its weights, and aims to inflate per-inference cost while
leaving accuracy unchanged. This is the conceptual precedent for
the present work.

Cin\`a et al.~\cite{cina2022sponge} subsequently showed that the
same harm model can be embedded into the model \emph{at training
time} via \emph{sponge poisoning}: an adversary who controls a
fraction of the training data biases the network toward producing
high-activity outputs at inference. te Lintelo et al.~\cite{telintelo2024skipsponge}
extended this direction with \emph{SkipSponge}, which achieves a
sponge-like effect by directly perturbing pretrained model weights
without retraining. Both works share Shumailov's harm model
(per-inference cost inflation under preservation) but operate on
the model itself rather than the input.

\subsection{Universal sponge attacks on dense ANNs}\label{sec:univ-sponge-ann}
Universal sponge attacks --- a single fixed perturbation that
inflates inference cost on \emph{any} input the deployed model
sees --- have, to our knowledge, only been demonstrated for specific Deep Neural Network (DNN) architectures. In particular, Phantom
Sponges~\cite{shapira2023phantom} crafts a universal adversarial
patch that exploits Non-Maximum Suppression (NMS) in object
detectors, inducing a flood of overlapping bounding boxes that
inflate NMS post-processing latency while leaving the detector's
forward-pass cost essentially unchanged. The authors explicitly
report that their initial attempt to apply Shumailov-style
universal sponge to YOLO's feedforward phase was unsuccessful.
Hong et al.~\cite{hong2021earlyexit} target multi-exit networks
with universal perturbations that prevent inputs from satisfying
intermediate confidence thresholds, forcing inference through all
layers; the exploited surface is the early-exit branch logic ---
an architectural feature --- rather than the dense forward pass.

In these works, the layer-wise compute cost is not increased; the attack finds inputs that increase post-processing cost ~\cite{shapira2023phantom} or makes the network exit later adding extra layer computations~\cite{hong2021earlyexit}. Event-driven SNNs
have a fundamental difference: their forward-pass cost is 
input-dependent and scales directly with spike count, creating
the attack surface this paper exploits.

\subsection{Input-space sponge attacks on SNNs}
To our knowledge, the only published input-space efficiency attack
that targets SNNs is \emph{SpikeAttack}~\cite{krithivasan2022spikeattack}.
The authors craft an additive, real-valued, input-specific 
perturbation under an $L_p$-norm budget that,
when added to a clean image and fed to a rate-coded SNN, increases
the spike count propagated through the network while preserving the
top-1 prediction. They report spike-inflation factors of approximately
1.7--2.5$\times$ at preservation rates above $\sim$95\%.

\textit{SpikeAttack} is evaluated on two image-classification benchmarks
(CIFAR-10 and ImageNet). In both cases the input is a real-valued
pixel-intensity image, and the actual input to the SNN is a rate-coded spike
train derived from those intensities. Neither benchmark is a
\emph{native event-based} dataset. Whether the rate-coded
input-space formulation transfers to the binary event-based
regime --- where the input alphabet is $\{0,1\}$ and an $L_p$-ball
constraint reduces to an $L_0$ flip count --- is therefore not
addressed by their work. The attack is also strictly per-sample;
universal sponge perturbations on SNNs are not considered.
Consequently, no direct numerical comparison with SpikeAttack is
possible: porting its continuous additive perturbation to a
$\{0,1\}$ input requires thresholding back to binary, which
destroys the gradient signal it relies on. The per-sample attack
we propose is, to our knowledge, the first sponge formulation
defined natively in the binary event-based regime.

\subsection{Parameter-space attacks on SNN energy}

A complementary line of work targets the network's \emph{parameters}
rather than its inputs. Yang et al.~\cite{11295756} (EOS)
achieve a sponge-like outcome by flipping a small number of bits in
the firing-threshold parameters via Row-Hammer, lowering the
threshold and causing neurons to fire more readily. EOS does not
craft an adversarial input; it modifies the deployed model at
inference time. We list it for completeness, as it shares the same
harm model (energy-oriented attack on an SNN) but addresses an
orthogonal attack surface.

\subsection{Position relative to prior work}

In summary, prior work covers per-sample sponge on rate-coded SNNs
~\cite{krithivasan2022spikeattack}, parameter-side sponge on SNNs \cite{11295756}, per-sample and
training-time sponge on dense ANNs ~\cite{shumailov2021sponge},~\cite{cina2022sponge},~\cite{telintelo2024skipsponge},
and architectural-quirk universal sponge on dense ANNs ~\cite{shapira2023phantom},~\cite{hong2021earlyexit}. 
Two gaps remain. First, sponge attacks in the
native binary event-based regime --- the regime in which DVS cameras
and silicon cochleae actually produce data, and in which SNNs make
operational sense as low-energy edge inference engines --- have not
been demonstrated. Second, universal sponge attacks against the
forward pass of any DNN remain open. This paper addresses both gaps for SNNs operating on native binary event-based inputs. 

For completeness, we note that sponge attacks are only one class of security threats against SNNs. Other attacks investigated in the literature include adversarial input generation \cite{BaSiRa18,BLHSS18,SPSLPR19,E-AMSA21,MPMMS21,LHDWLDLX23,MNKHMS20,
NSHM22,RaSt25b, RaSt26c}
, hardware Trojan \cite{VMAMS20,RKKAS25}, model stealing attacks \cite{BD-RADARS26} and ownership verification \cite{RaSt26d}, backdoor attacks \cite{AEPU24,10.1007/978-981-96-7005-5_17,10.1007/978-3-032-07884-1_1, 11391974}, fault injection attacks \cite{NLEKKG22}, side-channel attacks \cite{NRTKG23,GoDaSu24}, and privacy attacks \cite{aksu2025privacyfederatedlearningspiking}. 
\section{Threat Model}\label{sec:threat}
\subsection{Adversary capabilities}

The adversary has \emph{white-box} access to the deployed SNN
$f_\theta$: weights, neuron parameters, and architecture are known
to the attacker. This is a strong but standard assumption in
adversarial-attack evaluation. The adversary cannot
modify the weights $\theta$ at run time (parameter-space attacks
such as Row-Hammer-induced bit flips~\cite{11295756} are out of
scope here).

The adversary's harm model is to inflate the per-inference cost
(measured in synaptic operations --SynOps-- or, equivalently, in joules on
event-driven hardware) while preserving the network's predicted
class. Concretely, given a clean input
$\mathbf{x}_0$, the adversary
seeks to produce
$\mathbf{x}_{\mathrm{sponge}}$
such that:
\begin{enumerate}
\item The predicted class is preserved.
\item The SynOp count SynOps$(\mathbf{x}_{\mathrm{sponge}})$ is
substantially larger than SynOps$(\mathbf{x}_0)$.
\item The time-to-decision is comparable to (or below) the clean
input's, so the attack does not reveal itself through observable
latency.
\end{enumerate}
Operationally, the adversary is satisfied if the deployed device
keeps producing correct, on-time predictions while the battery
drains faster.

\subsection{Two attack regimes, two attack-surface assumptions}

The two regimes we study correspond to two qualitatively different
attacker capabilities at \emph{deployment time}.

\paragraph*{Per-sample regime}

The per-sample attack assumes the adversary can interpose a
significant amount of per-input compute between the sensor and the
SNN. For each clean spike train $\mathbf{x}_0$ the attacker runs a
gradient-based optimization algorithm to produce a custom
sponge $\mathbf{x}_{\mathrm{sponge}}$, then forwards
$\mathbf{x}_{\mathrm{sponge}}$ to the SNN in place of $\mathbf{x}_0$.
This produces large inflation but, as our timing measurements show, the per-input optimization
cost may be incompatible with the streaming inference
rate of a real device. We characterize the per-sample attack
primarily as a \emph{baseline upper bound} of the success of sponge attacks.

\paragraph*{Universal sponge regime}

In the universal sponge attack 
the adversary generates once and offline a single, fixed binary mask
$\mathbf{U}$ which is injected at the
sensor-pre\-processing stage and is XORed
with every clean output $\mathbf{x}_0$ of the sensor to produce the sponge input $\mathbf{x}_{\mathrm{sponge}}$ for the SNN. There is no
per-input compute requirement on the deployed system; the
constant-time XOR is the only attack-side cost at inference time.
This is realizable in practice via several mechanisms:
\begin{itemize}
\item a stuck-pixel pattern or fixed light source in an event
camera's field of view, producing a deterministic event overlay
that the attacker chose offline.
\item a compromised sensor driver or pre-processing firmware that
silently XORs the input tensor with a stored mask.
\item a fixed acoustic signal in the listening environment of a
silicon cochlea, producing a deterministic event pattern at the
sensor output.
\end{itemize}
We argue this is the realistic operational threat the SNN community
should worry about.

\begin{figure}[t]
\centering
\includegraphics[width=\columnwidth]{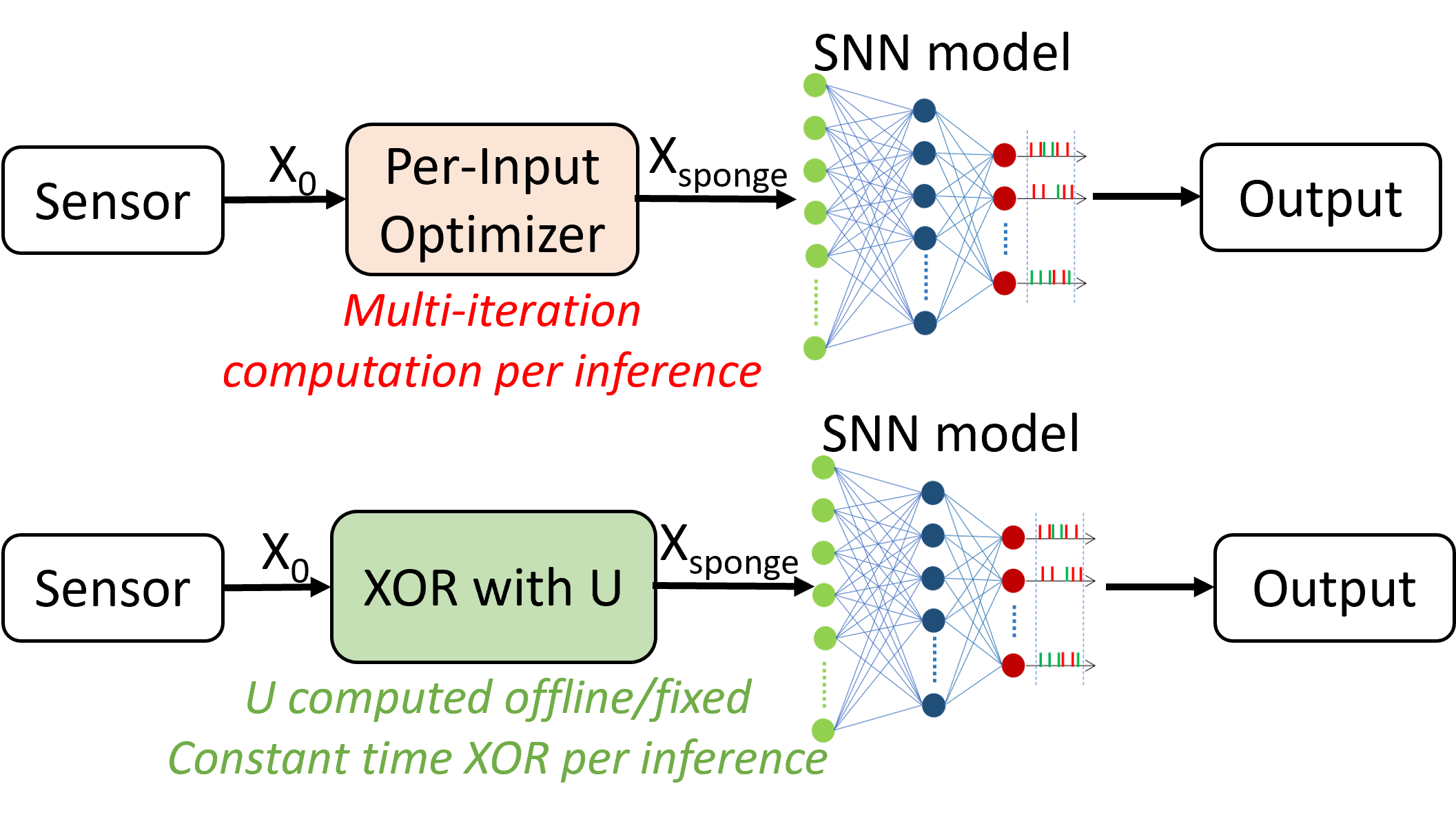}
\caption{Two attack regimes for sponge attacks on edge SNNs. Per-sample
(top). Universal sponge (bottom).}
\label{fig:threat_model}
\end{figure}

Fig.~\ref{fig:threat_model} contrasts the two regimes. The per-sample regime requires the attacker to perform a multi-iteration optimization for each sensor output to generate the sponge input for the SNN. In contrast, the universal regime requires only a one-time injection of a precomputed offline mask, $\mathbf{U}$, which is XORed with the sensor's output during operation.

\section{Preliminaries}\label{sec:notation}

\subsection{SNN notation}

A SNN is a discrete-time, event-driven function
approximator. Each neuron maintains an internal membrane potential
that increases with incoming spike events and decays between two consecutive input spikes, a mechanism referred to as leakage. When the membrane potential crosses
a pre-defined threshold, the neuron emits a spike to all postsynaptic neurons through its synaptic connections; otherwise it stays silent. For classification tasks, the network's prediction is determined by the output neuron with the highest spike count, where each output neuron represents a distinct class.

A clean input for SNNs is a binary spatio-temporal tensor

\begin{equation}
   \mathbf{x}_0 \in \{0,1\}^{C\times H\times W\times T}, 
\end{equation}

where $C$ is the number of input channels (e.g.\ $C\!=\!2$ on/off
polarities for an event camera), $H\!\times\! W$ is the spatial
resolution, and $T$ is the number of timesteps in the dataset sample.
The sponge counterpart $\mathbf{x}_{\mathrm{sponge}}$
is defined over the same domain.

An input produces per-layer spike
tensors $\mathbf{s}^{(\ell)}\!\in\!\{0,1\}^{n_\ell\times T}$ for
each layer $\ell\!\in\!\{1,\ldots,L\}$, where $n_\ell$ is the
number of neurons in layer $\ell$ and $L$ indexes the
output classifier layer. The first $L\!-\!1$ layers are
hidden; layer $L$ is composed of $n_L\!=\!K$
neurons corresponding to $K$ output classes. Let $s^{(\ell)}_{n,t}$ denote the output spike of neuron $n$ in layer $\ell$ at timestep $t$. 
For the output layer we use $s^{(L)}_{c,t}$, where $c$ denotes the class, $c=1,\cdots,K$. 
The SNN implements a function $f_\theta$ with fixed synaptic weights
$\theta$  

\begin{equation}
    \mathbf{s}^{(L)}=f_\theta \left ( \mathbf{x}\right ).
\end{equation}

\noindent The predicted class at timestep $r$ is

\begin{equation}
  y^\star_r(\mathbf{x}) \;=\; \arg\max_{c\in\{1,\ldots,K\}}\;
       \sum_{t=1}^r s^{(L)}_{c,t}(\mathbf{x}),
\end{equation}

\noindent with the final decision being $y^\star_T(\mathbf{x})$. The total spike count produced by the network when fed with
$\mathbf{x}$ is

\begin{equation}
\mathcal{S}(\mathbf{x}) \;=\;
  \sum_{\ell=1}^{L}\sum_{n=1}^{n_\ell}\sum_{t=1}^{T} s^{(\ell)}_{n,t}(\mathbf{x}).
\end{equation}

The hardware-relevant cost of an SNN inference is dominated by SynOps: each spike emitted by a neuron
contributes one SynOp per post-synaptic synapse, i.e.\ as many
SynOps as the neuron's fanout. Event-driven neuromorphic chips such as
Loihi~\cite{Loihi18} and TrueNorth~\cite{ASCA-IAMINDNTBBKMRJ15}
publish a per-SynOp energy figure (e.g.\ $23.6\,\mathrm{pJ}$ per
SynOp on Loihi-1), which directly converts the SynOp count of a
forward pass into joules per inference.

SynOps is expressed as
\begin{equation}
    \text{SynOps}(\mathbf{x})=\;
  \sum_{\ell=1}^{L}\sum_{n=1}^{n_\ell}\mathrm{fanout}(n)\sum_{t=1}^{T} s^{(\ell)}_{n,t}(\mathbf{x}),
\end{equation}

\noindent where $\mathrm{fanout}(n)$ is fanout of neuron $n$.

The estimated energy consumption for processing input $\mathbf{x}$ on neuromorphic processor chip $h$ is
\begin{equation}
    \mathcal{E}_h(\mathbf{x}) = \text{SynOps}(\mathbf{x})\cdot e_h,
\end{equation}
\noindent where $e_h$ is the chip's published per-SynOp energy, e.g., 
$e_{\text{Loihi-1}} = 23.6\,\mathrm{pJ}$ for the Loihi chip~\cite{Loihi18}.

\subsection{Attack metrics}

\subsubsection{Inflation in spike activity and energy consumption}

We quantify the impact of the sponge attack using the relative \emph{inflation in spike activity}, denoted by $R_S$, and \emph{inflation in energy consumption}, denoted by $R_E$

\begin{equation}
\begin{aligned}
  R_S(\mathbf{x}_{\mathrm{sponge}}) &= \frac{\mathcal{S}(\mathbf{x}_{\mathrm{sponge}})}{\mathcal{S}(\mathbf{x}_0)} \\
  R_E(\mathbf{x}_{\mathrm{sponge}}) &= \frac{\text{SynOps}(\mathbf{x}_{\mathrm{sponge}})}{\text{SynOps}(\mathbf{x}_0)}.
\end{aligned}
\end{equation}

\noindent Both these metrics are unitless. 

\subsubsection{Preservation}

The sponge input must maintain the predicted class of the clean input. We define \emph{preservation} $\Pi$ as the fraction of clean inputs $\mathbf{x}_0$ in the dataset $\mathcal{D}$ whose predicted class is
unchanged under their sponge counterpart $\mathbf{x}_{\mathrm{sponge}}$,

\begin{equation}\label{eq:preservation}
    \Pi = \frac{1}{|\mathcal{D}|}\sum_{\mathbf{x}_0 \in \mathcal{D}}
    \mathbbm{1}\!\left[\,y^\star_T(\mathbf{x}_{\mathrm{sponge}}) =
    y^\star_T(\mathbf{x}_0)\,\right] \;\in\; [0,1],
\end{equation}
where $\mathbf{x}_{\mathrm{sponge}}$ is the sponge counterpart of
$\mathbf{x}_0$ and $\mathbbm{1}[\cdot]$ is $1$ when the condition holds
and $0$ otherwise.
We accompany the preservation metric with the  Wilson $95\%$ binomial confidence interval. Preservation is an attacker-side metric: does the attack
keep the network's decisions identical?

\subsubsection{Sponge accuracy}

It is defined as the accuracy of the model over the sponge inputs where the expected labels are those of the corresponding clean inputs
\begin{equation}
    A = \frac{1}{|\mathcal{D}|}\sum_{\mathbf{x}_0 \in \mathcal{D}}
    \mathbbm{1}\!\left[\,y^\star_T(\mathbf{x}_{\mathrm{sponge}}) =
    y_{true}(\mathbf{x}_0)\,\right] \;\in\; [0,1],
\end{equation}

\noindent where $y_{true}(\mathbf{x}_0)$ denotes the true class of input $\mathbf{x}_0$. Sponge accuracy
is the defender-side metric: does the attack degrade the deployed
model's accuracy?

\subsubsection{Time-to-decision}

An attack that preserves the final prediction but causes a noticeable increase in inference latency can be detected by the user and is therefore not considered stealthy. To quantify this effect, we define the time-to-decision (TTD) as the earliest timestep at which the running prediction matches the final prediction
\begin{equation}
  \mathrm{TTD}(\mathbf{x}) = \min\bigl\{r\in\{1,\ldots,T\}: y^\star_r(\mathbf{x})= 
 y^\star_T(\mathbf{x})\bigr\}.
\end{equation}

We then define the difference in TTD as
\begin{equation}
  \mathrm{dTTD}(\mathbf{x}_{\mathrm{sponge}}) = \mathrm{TTD}(\mathbf{x}_{\mathrm{sponge}}) - \mathrm{TTD}(\mathbf{x}_0),
\end{equation}
 
\noindent To remain stealthy, the adversary aims to keep $\mathrm{dTTD}$ below a prescribed threshold $\gamma$, i.e., $\mathrm{dTTD}(\mathbf{x}_{\mathrm{sponge}})<\gamma$ (fast decision, more
energy, indistinguishable to the user).

\subsubsection{Input perturbation}

Because the input is binary, the natural perturbation budget is
$L_0$: the number of positions $(c,h,w,t)$ at which
$\mathbf{x}_{\mathrm{sponge}}$ differs from $\mathbf{x}_0$.

The two attack regimes share the harm model defined in
Section~\ref{sec:threat} but differ in the deployment-time
attacker capability they assume; the methodology makes this concrete.

\section{Proposed Techniques} \label{sec:method}
\subsection{Per-Sample Sponge Attack}

The per-sample attack crafts a custom sponge spike train
$\mathbf{x}_{\mathrm{sponge}}$ for each clean input $\mathbf{x}_0$ by
gradient optimization on a parameterised perturbation. We describe
the differentiable surrogate (which makes the binary input
optimisable), the objective with the rationale of each term, and the
optimization recipe.

\subsection*{Differentiable surrogate over a binary tensor}
Both the input and the gradient must be reconciled with the binary
alphabet $\{0,1\}$. We introduce a free parameter tensor of logits
$\mathbf{P}\!\in\!\mathbb{R}^{C\times H\times W\times T}$ of the same
shape as $\mathbf{x}_0$ and produce, at each optimization step, a
soft mask
\begin{equation}
    \mathbf{m} = \epsilon \cdot \mathrm{GumbelSoftmax}(\mathbf{P};\tau) \in \mathbb{R}^{C\times H\times W\times T},
\end{equation}
where $\epsilon$ is a fixed scale factor. The Gumbel-Softmax
ensures $\mathbf{m}\!\in\![0,\epsilon]^{C\times H\times W\times T}$
and is controlled by the temperature parameter $\tau$. Higher $\tau$ 
produces softer masks, while lower $\tau$ produces masks closer to 
binary.

We then apply $\mathbf{m}$ as a \emph{symmetric flip rule}: at each
position, when applied on a clean-zero entry tries to push it up to a
spike, and on a clean-one entry tries to suppress it:
\begin{equation}
     \tilde{\mathbf{x}} =
   \mathbf{x}_0 + (\mathbf{1}-\mathbf{x}_0)\odot\mathbf{m}
                - \mathbf{x}_0\odot\mathbf{m}. 
\end{equation}

A straight-through estimator binarizes $\tilde{\mathbf{x}}$ at
threshold $0.5$ during the forward pass and passes the upstream
gradient through unchanged during the backward pass:
\begin{equation}
    \mathbf{x}_{\mathrm{sponge}} = \mathrm{STE}\!\bigl(\tilde{\mathbf{x}}\bigr).
\end{equation}
The result is a binary $\mathbf{x}_{\mathrm{sponge}}$ on which the
network forward pass operates exactly as it would on the clean
tensor, while gradients can flow back to $\mathbf{P}$ through the soft
mask. The forward pass through the victim network exposes the
per-layer hidden spike tensors and the output spike tensor $\{\mathbf{s}^{(\ell)}\}_{\ell=1}^{L}$ used
by the objectives below.

\subsection*{Objective functions}
After the forward pass, we compute three objective functions
combining a sponge term, a preservation term, and a
variance regularizer:

\subsubsection*{Sponge term (rationale: drive every neuron to saturation)}
We use a piecewise-linear hinge that drives each neuron $n$ in each layer $\ell$ to spike at every time bin, pushing it
toward its maximum possible spike count $T$:
\begin{equation}\label{eq:force}
  \mathcal{L}_{\mathrm{force}} \;=\;
    \sum_{\ell=1}^{\mathcal{L}}\sum_{n=1}^{n_\ell}
    \max\!\bigl(0,\; T - \!\sum_{t=1}^T s^{(\ell)}_{n,t}(\mathbf{x}_{\mathrm{sponge}})\bigr).
\end{equation}
\emph{Why this form}: the hinge is zero
when a neuron already fires $T$ times (saturated; no more inflation
possible); positive otherwise. Its gradient is constant in the
non-saturated regime, so each under-spiking neuron contributes
proportionally to the descent direction, and the total cost grows
linearly with the per-neuron deficit summed across the network.

\subsubsection*{Preservation term (rationale: actively push spikes through the clean class)}
Let $z^{(L)}_{y^\star}\!=\!\sum_{t=1}^T s^{(L)}_{y^\star,t}(\mathbf{x}_{\mathrm{sponge}})$
be the output-layer spike count of the clean predicted class
$y^\star=y^{\star}_{T}(x_0)$ on the sponge input. We use the reciprocal surrogate
\begin{equation}\label{eq:pres}
  \mathcal{L}_{\mathrm{pres}} \;=\;
    \frac{1}{z^{(L)}_{y^\star} + \varepsilon_{\mathrm{div}}},
\end{equation}
with $\varepsilon_{\mathrm{div}}\!=\!10^{-7}$. 
\emph{Why this form}: The reciprocal surrogate is active
\emph{throughout} optimization. Concretely,
$\partial \mathcal{L}_{\mathrm{pres}} / \partial z^{(L)}_{y^\star} = -1/(z^{(L)}_{y^\star} + \varepsilon_{\mathrm{div}})^2$
is finite and strictly negative everywhere on the admissible
domain $z^{(L)}_{y^\star}\!\ge\!0$ (the $\varepsilon_{\mathrm{div}}$
guards against the singularity at $z^{(L)}_{y^\star}\!=\!0$), so
the optimizer is pushed to grow the clean-class output spike count
throughout optimization, not only when preservation is about to fail. We found this necessary on
event-based regimes where the clean prediction margin is fragile.

\subsubsection*{Variance regularizer (rationale: soft $L_0$ budget for stealth)}
\begin{equation}
    \mathcal{L}_{\mathrm{var}} = \mathrm{Var}(\mathbf{x}_0 - \mathbf{x}_{\mathrm{sponge}})
\end{equation}
acts as a soft $L_0$ penalty: on sparse binary inputs, minimizing
the variance of the perturbation is to first order minimizing
the number of flipped positions, which keeps the sponge's
footprint at the sensor small and harder for an
input-distribution monitor to flag. It functions
as a soft tiebreaker rather than a hard budget.

\subsubsection*{Total objective}
The total objective $\mathcal{L}$ is a weighted sum of the three terms:
\begin{equation}
  \mathcal{L} \;=\;
   w_{\mathrm{var}}\,\mathcal{L}_{\mathrm{var}}
   + w_{\mathrm{f}}\,\mathcal{L}_{\mathrm{force}}
   + w_{\mathrm{p}}\,\mathcal{L}_{\mathrm{pres}}.
   \label{eq:total_loss}
\end{equation}

\subsection*{Optimization recipe}
We optimize $\mathbf{P}$ with Adam~\cite{Adam2017optimization} for a fixed
budget of $N_{\mathrm{iter}}$ iterations, with a plateau-based learning-rate
schedule on the total loss and a cyclical schedule for the Gumbel
temperature $\tau$ (Algorithm~\ref{alg:adv}). The update at each iteration is
\begin{equation}
    \mathbf{P} \leftarrow \mathbf{P} - lr \cdot \nabla_{\mathbf{P}} \mathcal{L}.
    \label{eq:adam_step}
    \vspace{-0.1cm}
\end{equation}
We track the running best $\mathbf{x}_{\mathrm{sponge}}$ across iterations
and return the saved best as the attack output.

\begin{algorithm}[t]
\caption{Per-sample sponge attack}\label{alg:adv}
\KwIn{clean $\mathbf{x}_0$, network $f_\theta$, weights
$w_{\mathrm{var}}, w_{\mathrm{f}}, w_{\mathrm{p}}$, scale $\epsilon$,
iterations $N_{\mathrm{iter}}$}
\KwOut{$\mathbf{x}_{\mathrm{sponge}}\in\{0,1\}^{C\times H\times W\times T}$}
$y^\star \leftarrow$ clean predicted class of $\mathbf{x}_0$\;
$\mathbf{P}\sim\mathcal{N}(0,1)$\; 
$\mathbf{x}_{\mathrm{sponge}} \leftarrow \mathbf{0}$;
$\mathcal{L}^\star \leftarrow +\infty$\;
\For{$it = 1$ \KwTo $N_{\mathrm{iter}}$}{
  $\tau \leftarrow$ next temperature in schedule\;
  $\mathbf{m} \leftarrow \epsilon\cdot\mathrm{GumbelSoftmax}(\mathbf{P};\tau)$\;
  $\tilde{\mathbf{x}} \leftarrow \mathbf{x}_0 + (\mathbf{1}-\mathbf{x}_0)\odot\mathbf{m} - \mathbf{x}_0\odot\mathbf{m}$\;
  $\mathbf{x}_{\mathrm{sponge}} \leftarrow \mathrm{STE}(\tilde{\mathbf{x}})$\;
  forward $f_\theta(\mathbf{x}_{\mathrm{sponge}})$ and capture per-layer spikes\;
  evaluate $\mathcal{L}$ from \eqref{eq:total_loss}\;
  \If{$\mathcal{L}<\mathcal{L}^\star$}{
    $\mathbf{x}_{\mathrm{sponge}}^\star\leftarrow\mathbf{x}_{\mathrm{sponge}}$;
    $\mathcal{L}^\star\leftarrow\mathcal{L}$\;
  }
  Adam step on $\mathbf{P}$ from \eqref{eq:adam_step}\;
}
\Return $\mathbf{x}_{\mathrm{sponge}}^\star$\;
\end{algorithm}

\subsection{Universal Sponge Attack}\label{sec:univ}
The per-sample attack is computed
\emph{per input}: every clean spike train passes through a
multi-iteration Adam loop before the corresponding
$\mathbf{x}_{\mathrm{sponge}}$ is forwarded to the SNN. Operationally
this requires the adversary to interpose a latency-prohibitive
optimization between each sensor reading and the deployed
network. In a streaming-edge deployment that is unrealistic.

The universal sponge attack relaxes this assumption. We construct
once, offline, a single fixed binary perturbation
$\mathbf{U}\in\{0,1\}^{C\times H\times W\times T}$ from a small
training-set subset, and apply it to every subsequent input by
elementwise XOR. Construction takes about a minute; per-inference
attack cost is a constant-time XOR with a fixed mask and is
effectively free at the device level. This is the regime in which a
sensor-side overlay attack (a stuck-pixel pattern, a fixed light
source in a DVS camera's field of view, a compromised sensor
driver, or a stationary acoustic signal in a silicon cochlea's
listening environment) becomes operational. We argue this is the
realistic threat the SNN community should worry about.

\subsection*{Construction}

The construction is a single forward+backward pass over a stratified
subset of the \emph{training} split. Let
$\mathcal{D}_{\mathrm{train}} = \{\mathbf{x}^{(i)}\}_{i=1}^M$ be the
construction set, where $M$ is small. We use $10$ samples per class
in this paper, giving $M=100$ on NMNIST ($10$ classes), $M=200$ on
SHD ($20$ classes), and $M=110$ on IBM DVS Gesture ($11$ classes).
For each sample we forward $\mathbf{x}^{(i)}$ through $f_\theta$,
compute the sponge term $\mathcal{L}_{\mathrm{force}}$ from
\eqref{eq:force}, and accumulate the per-input gradient with respect
to the input:
\begin{equation}\label{eq:grad_accum}
  \mathbf{g} \;=\; \frac{1}{M}\sum_{i=1}^{M}
    \nabla_{\mathbf{x}}\mathcal{L}_{\mathrm{force}}(f_\theta(\mathbf{x}^{(i)})).
\end{equation}
The accumulated gradient $\mathbf{g}$ is the direction in which
flipping a position would on average reduce $\mathcal{L}_{\mathrm{force}}$
across the construction set.

\subsubsection*{Why we omit the preservation term during construction}
Unlike the per-sample objective \eqref{eq:total_loss}, the construction
objective above contains \emph{no preservation term}. The reason is
sample-specific gradient cancellation: the clean predicted class
$y^\star$ varies sample to sample, so the direction of the
preservation gradient at any given input position has a different
sign for different samples and the average is uninformative. We
therefore optimize for spike inflation only and \emph{measure}
preservation post-hoc as an outcome of the construction.

\subsection*{Binarization}

We first rescale $\mathbf{g}$ to unit standard deviation,
\begin{equation}
\tilde{\mathbf{g}} \;=\; \mathbf{g}/\bigl(\mathrm{std}(\mathbf{g}) + \epsilon_{\mathrm{div}}\bigr),
\end{equation}
so the slope hyperparameter $\varepsilon$ in the next step has 
comparable effect across datasets regardless of raw gradient
magnitude. The sign pattern and relative ordering of entries 
in $\mathbf{g}$ are preserved.

We then form the per-position flip logit
\begin{equation}
  \ell_g \;=\; -\varepsilon\,\tilde{\mathbf{g}}+ b,
\end{equation}
a uniform baseline $b$ plus a per-position gradient bonus
$-\varepsilon\,\tilde{\mathbf{g}}$ (the minus makes
attacker-favourable positions $\tilde g\!<\!0$ score high). The
slope $\varepsilon$ controls how strongly the gradient steers the
choice; the bias $b$ sets the default flip probability, with negative $b$ keeping perturbation sparse.
The pair $(\varepsilon, b)$ gives smooth
control over the realized perturbation. We use a single
recommended operating point per dataset (Section~\ref{sec:setup})
rather than a sweep.
A Gumbel-Softmax at temperature $\tau$
then turns $\boldsymbol{\ell_g}$ into a soft flip probability,
\begin{equation}
  \mathbf{p_{universal}} \;=\; \mathrm{GumbelSoftmax}\bigl(\ell_g;\,\tau\bigr).
\end{equation}
Finally, we threshold the soft sample at $0.5$:
\begin{equation}\label{eq:univ_bern}
  \mathbf{U} \;=\; \mathbbm{1}\!\bigl[\mathbf{p_{universal}}> 0.5\bigr].
\end{equation}
The result $\mathbf{U}\!\in\!\{0,1\}^{C\times H\times W\times T}$
is the binary universal sponge mask which can be applied to a clean input by elementwise XOR ($\oplus$ denotes elementwise XOR
on $\{0,1\}$-valued tensors):
\begin{equation}\label{eq:univ_apply}
  \mathbf{x}_{\mathrm{sponge}} \;=\; \mathbf{x}_0 \oplus \mathbf{U}.
\end{equation}
Construction is one-shot and the application is constant-time XOR. There
is no Adam loop, no per-input optimization.
A wearable that runs an SHD-style keyword spotter would pay the universal sponge's
per-inference cost as the cost of a single bitwise XOR over the
input tensor, which on modern hardware is microseconds and
undetectable.
\section{Experimental Setup}\label{sec:setup}
We attack three SLAYER-trained~\cite{shor18} SNNs that span
three orders of magnitude in input dimensionality and cover both
visual (event-camera) and auditory (silicon-cochlea) sensing
modalities.

\begin{figure}[htbp]
\centering
\begin{subfigure}{0.4\textwidth}
    \includegraphics[width=\linewidth]{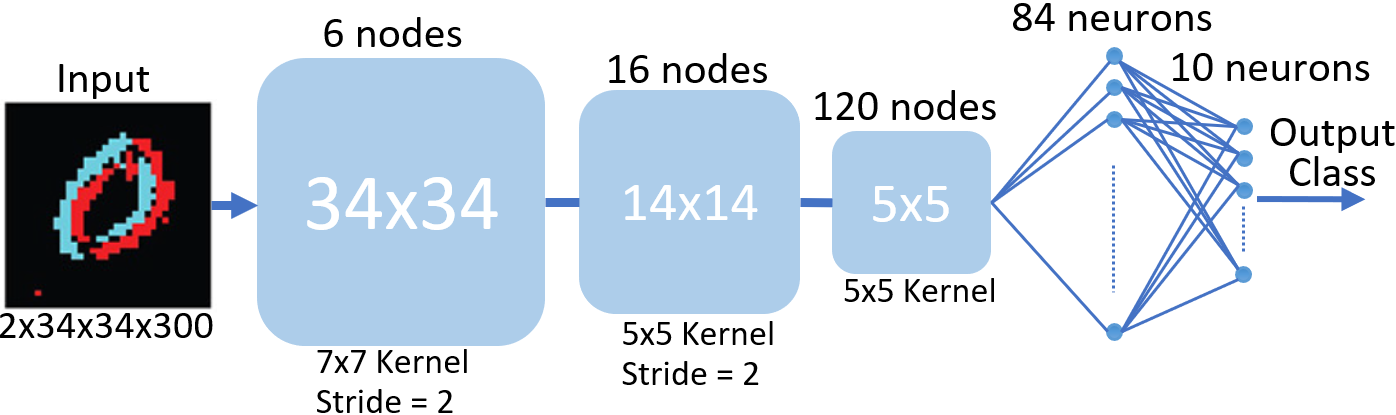}
    \caption{N-MNIST Architecture}
    \vspace{0.2cm}
\end{subfigure}
\begin{subfigure}{0.47\textwidth}
    \includegraphics[width=\linewidth]{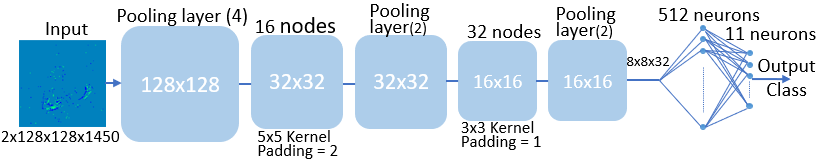}
    \caption{IBM DVS Gesture Architecture}
    \vspace{0.2cm}
\end{subfigure}
\begin{subfigure}{0.45\textwidth}
    \includegraphics[width=\linewidth]{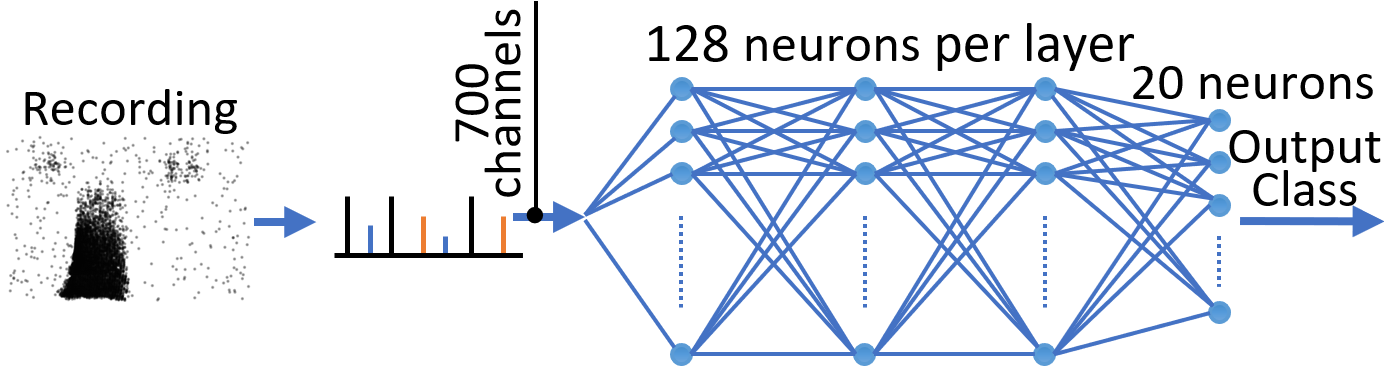}
    \caption{SHD Architecture}
\end{subfigure}
\caption{Spiking Neural Network Architectures used for the three case studies.}
\label{fig:architectures}
\vspace{-0.2cm}
\end{figure}

\begin{itemize}
\item \textbf{NMNIST.} The 10-class digit dataset of Orchard
et~al.~\cite{OJCT15}, presented as binary polarity events from a
saccading DVS over still MNIST images. It serves as a standard
benchmark for event-based vision. Each saccade scan is 
$300$\,ms long and is captured at the camera's native
$1$\,ms resolution into $T\!=\!300$ time steps over $C\!=\!2$
polarity channels and a $34\!\times\!34$ spatial grid,
giving a per-sample shape $(2,34,34,300)$.
\item \textbf{SHD (Spiking Heidelberg Digits).} The 20-class
spoken-digit dataset of Cramer et~al.~\cite{heidelberg}, produced
by a software silicon-cochlea model, demonstrating the cross-modal
applicability of our approach (vision and audio). Each utterance
is approximately $1\,000$\,ms long; we bin the raw cochlear spike
output at a $4$\,ms resolution into $T\!=\!250$ time steps over
$C\!=\!700$ frequency channels, giving a per-sample shape
$(700,250)$.
\item \textbf{IBM DVS Gesture.} The 11-class gesture dataset of
Amir et~al.~\cite{ATBM17}, recorded directly with an iniLabs DAVIS
event camera. It is used to evaluate scalability to
higher-dimensional spatio-temporal inputs. Each gesture is 
$1\,450$\,ms long and is captured at the camera's
native $1$\,ms resolution into $T\!=\!1\,450$ time steps over
$C\!=\!2$ polarity channels and a $128\!\times\!128$
spatial grid, giving a per-sample shape $(2,128,128,1\,450)$.
\end{itemize}
For each dataset, we train task-appropriate SNN
architectures using the SLAYER framework \cite{shor18}. The network architectures used
in each case study are illustrated in Fig.~\ref{fig:architectures}, and the per-model dataset splits and
baseline test accuracies are summarized in
Table~\ref{tab:victims}.

\begin{table}[t]
\centering
\caption{Summary of the three victim networks.}
\label{tab:victims}
\resizebox{\columnwidth}{!}{%
\begin{tabular}{lccc}
\toprule
 & NMNIST & SHD & IBM DVS \\
\midrule
Input shape & $(2,34,34,300)$ & $(700,250)$ & $(2,128,128,1\,450)$ \\
Input positions & $693\,600$ & $175\,000$ & $47\,513\,600$ \\
Train-set size & $60\,000$ & $8\,156$ & $1\,176$ \\
Test-set size & $10\,000$ & $2\,264$ & $264$ \\
Baseline accuracy & $98.2\%$ & $76.15\%$ & $86.7\%$ \\
\bottomrule
\end{tabular}}
\end{table}

\subsection*{Per-sample attack configuration}
For each dataset we evaluate the per-sample attack on the full
test set. We do not impose a hard $L_0$ projection; the realized
$L_0$ is determined by the optimization and reported per-sample.
The soft-mask scale $\epsilon$ from Section~\ref{sec:method} is
tuned per dataset on a small held-out subset, so that the soft
mass is large enough to drive the gradient through the
straight-through estimator without saturating the post-binarization
flip rate.

The objective weights of Eq.~(\ref{eq:total_loss})
are dataset-specific because the clean prediction margin and the
typical magnitudes of the three terms differ across datasets; we
tune them per dataset on the same held-out subset to balance
inflation and preservation throughout the optimization.

\subsection*{Universal sponge configuration}

The universal sponge construction set is a stratified subset of the
\emph{training} split: $10$ samples per class on each dataset
(unique deterministic indices, fixed seed), giving $M=100/200/110$ on
NMNIST/SHD/IBM DVS Gesture. For each dataset we use a single fixed
operating point $(\varepsilon, b, \tau)$, reported in
Table~\ref{tab:univ_op}, chosen to balance spike inflation against
preservation and then frozen across all reported experiments. The
resulting mask $\mathbf{U}$ is applied by elementwise XOR to every
input, and the attack is evaluated on the \emph{full} test set of
each dataset.

\begin{table}[t]
\centering
\caption{Recommended universal-sponge operating points. The values reported
in Section~\ref{sec:results} use these settings.}
\label{tab:univ_op}
\small
\begin{tabular}{lcccc}
\toprule
SNN model & Construction set & $\varepsilon$ & $b$ & $\tau$ \\
\midrule
NMNIST          & $100$ ($10/\text{class}$) & $1.0$ & $-6$  & $1.0$ \\
SHD             & $200$ ($10/\text{class}$) & $1.0$ & $-8$  & $1.0$ \\
IBM DVS Gesture & $110$ ($10/\text{class}$) & $1.6$ & $-13$ & $1.0$ \\
\bottomrule
\end{tabular}
\end{table}

\section{Results} \label{sec:results}
\subsection{Per-sample sponge attack}\label{sec:res-persample}

Table~\ref{tab:per_sample_main} reports the per-sample attack on
the \emph{full} test set of each dataset. Spike inflation is
substantial on every dataset ($1.80$ to $2.46\times$), preservation
is uniformly very high ($\Pi\!\geq\!0.9848$ with Wilson lower
bounds at least $0.96$), accuracy degradation is small to negligible
($-0.09$ to $-1.5$ percentage points), and time-to-decision
shifts are stealth-positive or essentially neutral on every
dataset.

\begin{table*}[t]
\centering
\caption{Per-sample sponge attack, full-test-set evaluation.}
\label{tab:per_sample_main}
\begin{tabular}{lccccccccccc}
\toprule
SNN model & samples & $R_S$ & $R_E$ & $\Pi$ & $\Pi$ Wilson 95\% CI & clean\_acc & sponge\_acc (A) & $\Delta$acc & $L_0$ & dTTD \\
\midrule
NMNIST          & $10\,000$ & $1.87$ & $1.50$ & $0.9946$ & $[0.993,\,0.996]$ & $0.982$& $0.977$ &$-0.5$ pp& $5\,494$  & $-31.7$ \\
SHD             & $2\,264$  & $1.80$ & $1.61$ & $0.9991$ & $[0.997,\,1.000]$ & $0.7615$& $0.7606$ &$-0.09$ pp& $2\,167$  & $-26.4$ \\
IBM DVS Gesture & $264$     & $2.46$ & $2.60$ & $0.9848$ & $[0.962,\,0.994]$ & $0.867$ & $0.852$ &$-1.5$ pp& $54\,569$ & $+5.7$ \\
\bottomrule
\end{tabular}
\end{table*}

\subsection*{Per-dataset narrative}

\subsubsection*{\textbf{NMNIST}} Spike count inflates by $1.87\times$, SynOps by
$1.50\times$, with preservation at $\Pi\!=\!0.9946$ on $10\,000$
test samples (Wilson 95\% CI $[0.993, 0.996]$). Classification
accuracy drops only $0.5$ percentage points ($98.2\%\to 97.7\%$)
because the attack changes very few predictions overall
($\sim\!99\%$ of its originally-correct
predictions remain correct under the attack). Realized $L_0\!\approx\!5\,494$
flips per sample, $\sim\!0.8\%$ of the input. dTTD is strongly
stealth-positive ($-31.7$ time bins, $-10.6\%$ of $T$).

\subsubsection*{\textbf{SHD}} Spike count inflates by $1.80\times$ at near-perfect
preservation ($\Pi\!=\!0.9991$, only $2$ of $2\,264$ samples
flipped; CI $[0.997, 1.000]$). $L_0$ is the smallest across the
three datasets ($\sim\!2\,167$ flips per sample). The accuracy
drop is essentially zero ($-0.09$ pp). dTTD $-26.4$ bins
(stealth-positive).

\subsubsection*{\textbf{IBM DVS Gesture}} The largest victim produces the strongest
inflation: $R_S\!=\!2.46$, $R_E\!=\!2.60$, with $\Pi\!=\!0.9848$
on the full $264$-sample test set (CI $[0.962, 0.994]$).
Sponge accuracy drops by $1.5$ percentage points
($86.7\%\to85.2\%$). The realized $L_0\!\approx\!54\,569$
is just $\sim\!0.11\%$ of IBM's $4.75\!\cdot\!10^7$-position input.
dTTD is small and positive ($+5.7/1\,450$ bins, $0.4\%$ of the
inference window).

\begin{figure}[t]
\centering
\includegraphics[width=\columnwidth]{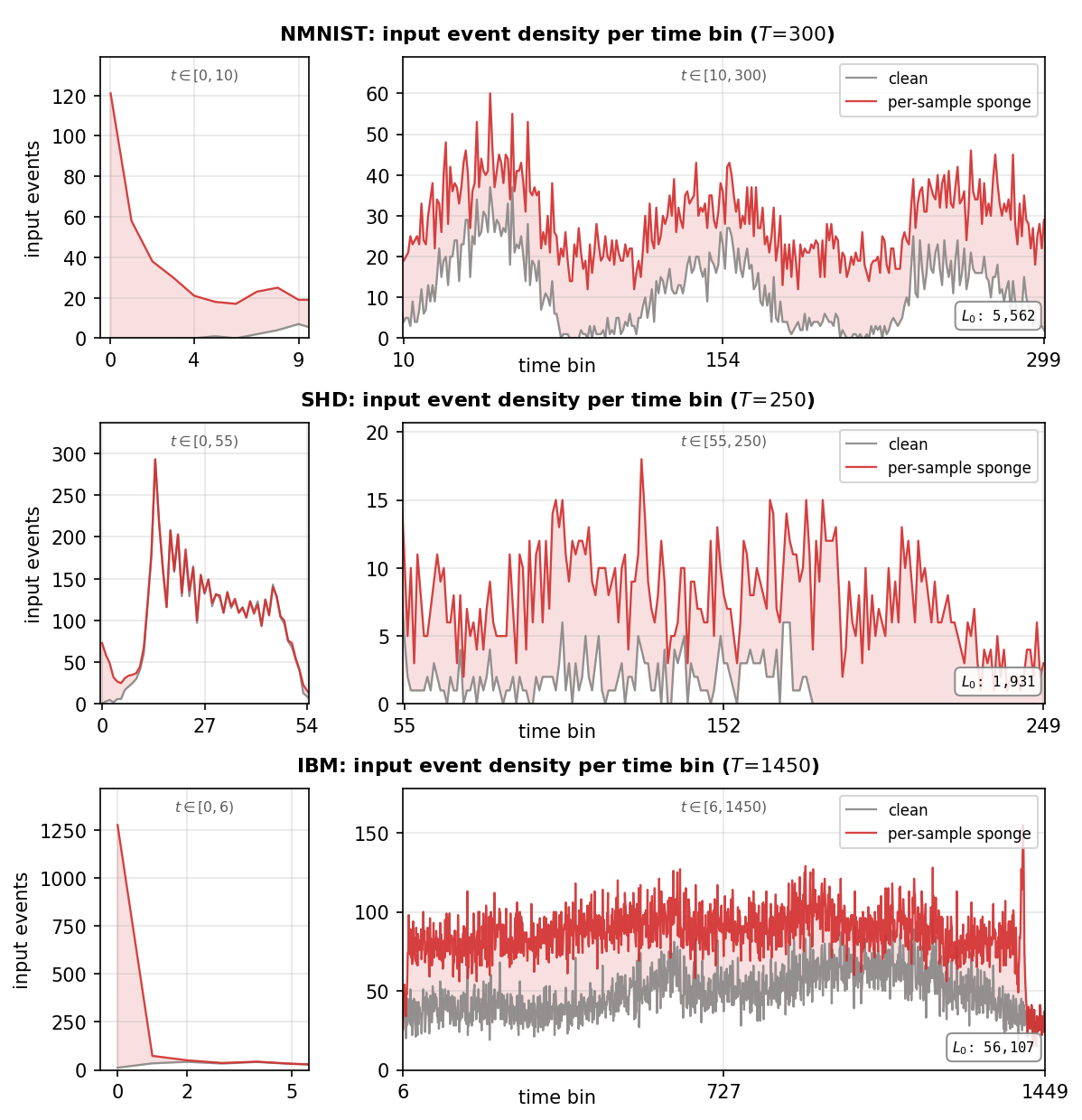}
\caption{Per-sample sponge: input-event count per time bin, clean
vs.\ sponge, representative sample per dataset.}
\label{fig:splitaxis_persample}
\end{figure}

\begin{figure}[t]
\centering
\includegraphics[width=\columnwidth]{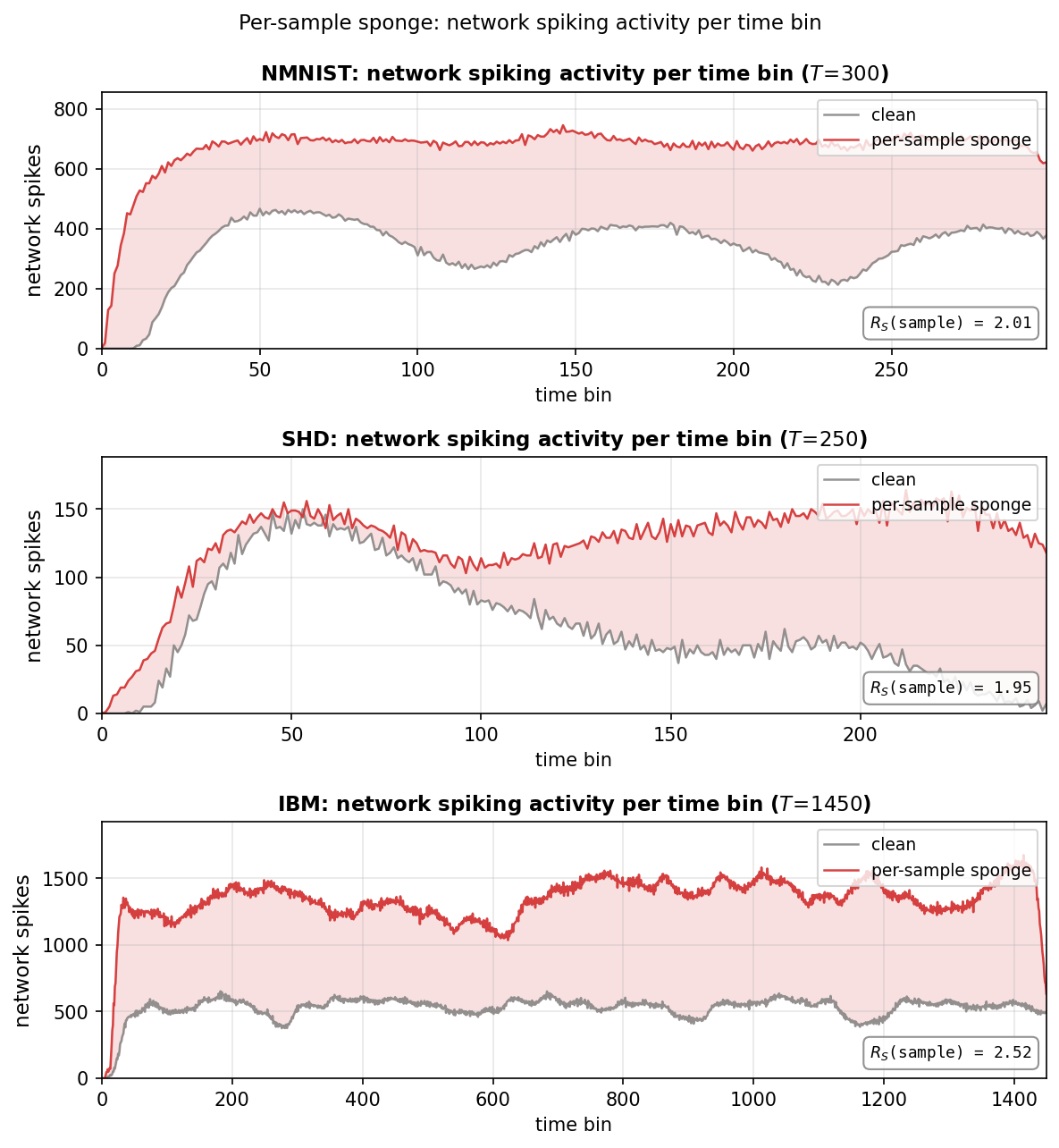}
\caption{Per-sample sponge: total network spikes per
time bin, clean vs.\ sponge, for a representative sample per dataset.}
\label{fig:netact_persample}
\end{figure}

\begin{figure}[t]
\centering
\includegraphics[width=\columnwidth]{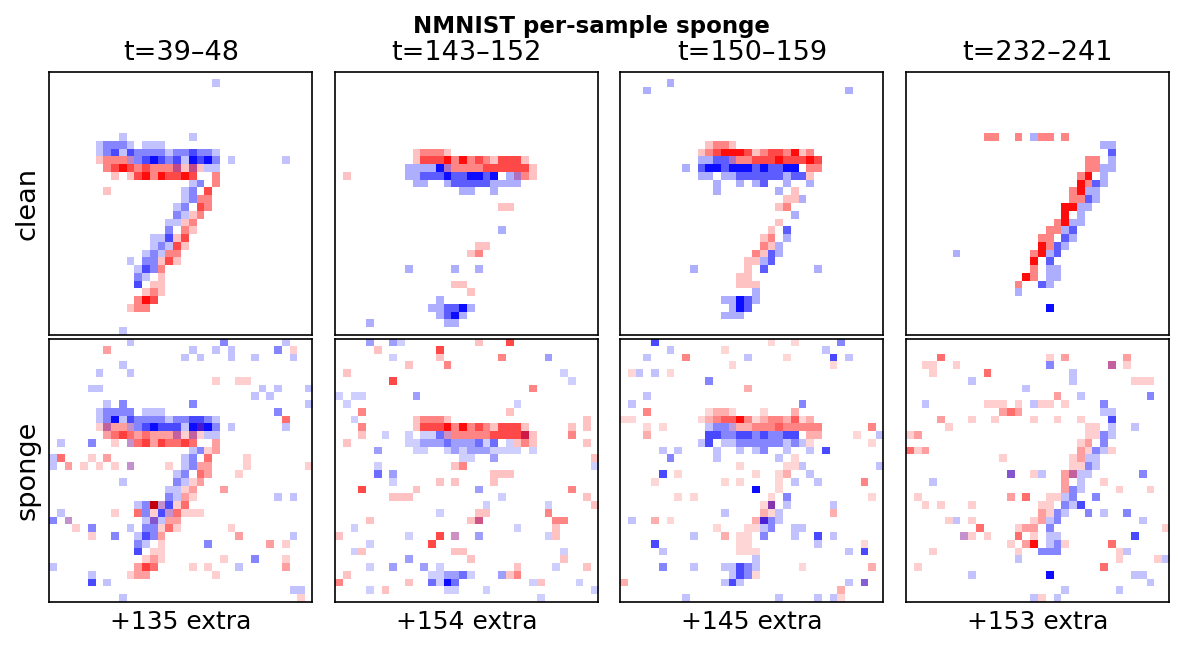}
\caption{NMNIST per-sample sponge: clean (top) vs.\ sponged
(bottom), four $10$-bin windows (blue\,=\,OFF, red\,=\,ON).}
\label{fig:lowpert_nmnist_persample}
\end{figure}

\begin{figure}[t]
\centering
\includegraphics[width=\columnwidth]{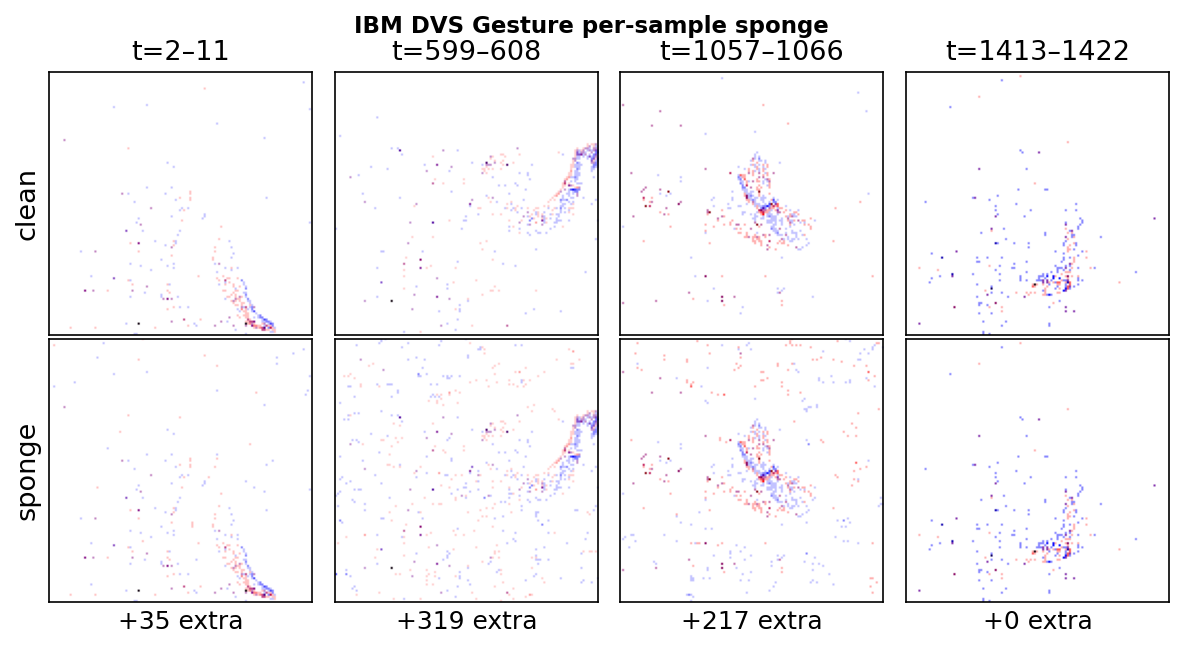}
\caption{IBM DVS Gesture per-sample sponge; construction as
Figure~\ref{fig:lowpert_nmnist_persample}.}
\label{fig:lowpert_ibm_persample}
\end{figure}

Figure~\ref{fig:splitaxis_persample} shows how the per-sample
attack distributes its injected events along the inference window. The
time axis is split so that the spike inflation is well-observed across $T$. Figure~\ref{fig:netact_persample} shows the network's internal
spiking activity per time bin under the clean and the sponged
input: the attack drives a sustained, network-wide increase in
spike activity across the whole inference window---the inflation
that translates directly into SynOps and energy.
Figures~\ref{fig:lowpert_nmnist_persample} and
\ref{fig:lowpert_ibm_persample} show the clean and sponged input
side by side for a representative NMNIST and IBM sample; SHD has no
spatial structure to visualize and is omitted from this view.

\subsection{Universal sponge attack}\label{sec:res-univ}

Table~\ref{tab:univ_main} reports the universal sponge at the
recommended operating point of Table~\ref{tab:univ_op}, validated
on the \emph{full} test set of each dataset. We list both
$\Pi$ (agreement with the clean prediction, the conservative
metric used in the sponge-attack literature) and sponge
accuracy (agreement with the true label, the defender's metric). We also report the Wilson $95\%$ CI on $\Pi$. The confidence intervals get wider as the test set gets smaller. They are tight at NMNIST scale
($\pm 0.005$), comfortable at SHD scale ($\pm 0.02$), and looser
but still informative at IBM scale ($\pm 0.04$).

\begin{table*}[t]
\centering
\caption{Universal sponge attack at the recommended operating point
per dataset, full-test-set evaluation.}
\label{tab:univ_main}
\begin{tabular}{lcccccccccc}
\toprule
SNN model & samples & $|\mathbf{U}|_0$ & $R_S$ & $R_E$ & $\Pi$ & $\Pi$ Wilson 95\% CI & clean\_acc & sponge\_acc (A) & $\Delta$acc & dTTD \\
\midrule
NMNIST          & $10\,000$ & $2\,936$ & $1.39$ & $1.19$ & $0.91$ & $[0.8993,\,0.9108]$ & $0.982$ & $0.902$ & $-8.0$ pp & $+41.55$\\
SHD             & $2\,264$  & $304$    & $1.33$ & $1.24$ & $0.73$ & $[0.7128,\,0.7493]$ & $0.7615$ & $0.658$ & $-10.35$ pp &$-6.6$\\
IBM DVS Gesture & $264$     & $30\,465$ & $1.05$ & $1.09$ & $0.89$ & $[0.8424,\,0.9192]$ & $0.867$ & $0.811$ & $-5.6$ pp &$+11.73$\\
\bottomrule
\end{tabular}
\end{table*}

\subsubsection*{\textbf{NMNIST}} A single $2\,936$-flip universal sponge
($0.42\%$ of the input tensor) inflates spike count by $1.39\times$
across $10\,000$ unseen test samples. Sponge accuracy drops
from $98.2\%$ to $90.2\%$ (an $8.0$ percentage-point drop), with
$91\%$ of predictions matching the clean prediction.

\subsubsection*{\textbf{SHD}} A $304$-flip universal sponge inflates spike
count by $1.33\times$. Sponge accuracy drops from $76.15\%$ to
$65.8\%$ (a $10.35$ pp drop). dTTD on SHD is negative
($-6.6$ time bins), i.e.\ the universal sponge causes the network
to reach its decision $\sim\!2.5\%$ \emph{earlier} than the clean
input (stealth-positive).

\subsubsection*{\textbf{IBM DVS Gesture}} A $30\,465$-flip universal sponge
inflates spike count by $1.05\times$ and SynOps by $1.09\times$,
with $\Pi\!=\!0.89$ and sponge accuracy dropping from $86.7\%$
to $81.1\%$ ($-5.6$ pp). The inflation \emph{ratio} on IBM is
modest because of the input's high dimensionality, but the
absolute energy delta (Section~\ref{sec:results-energy}) is the
largest of the three datasets.

\begin{figure}[t]
\centering
\includegraphics[width=\columnwidth]{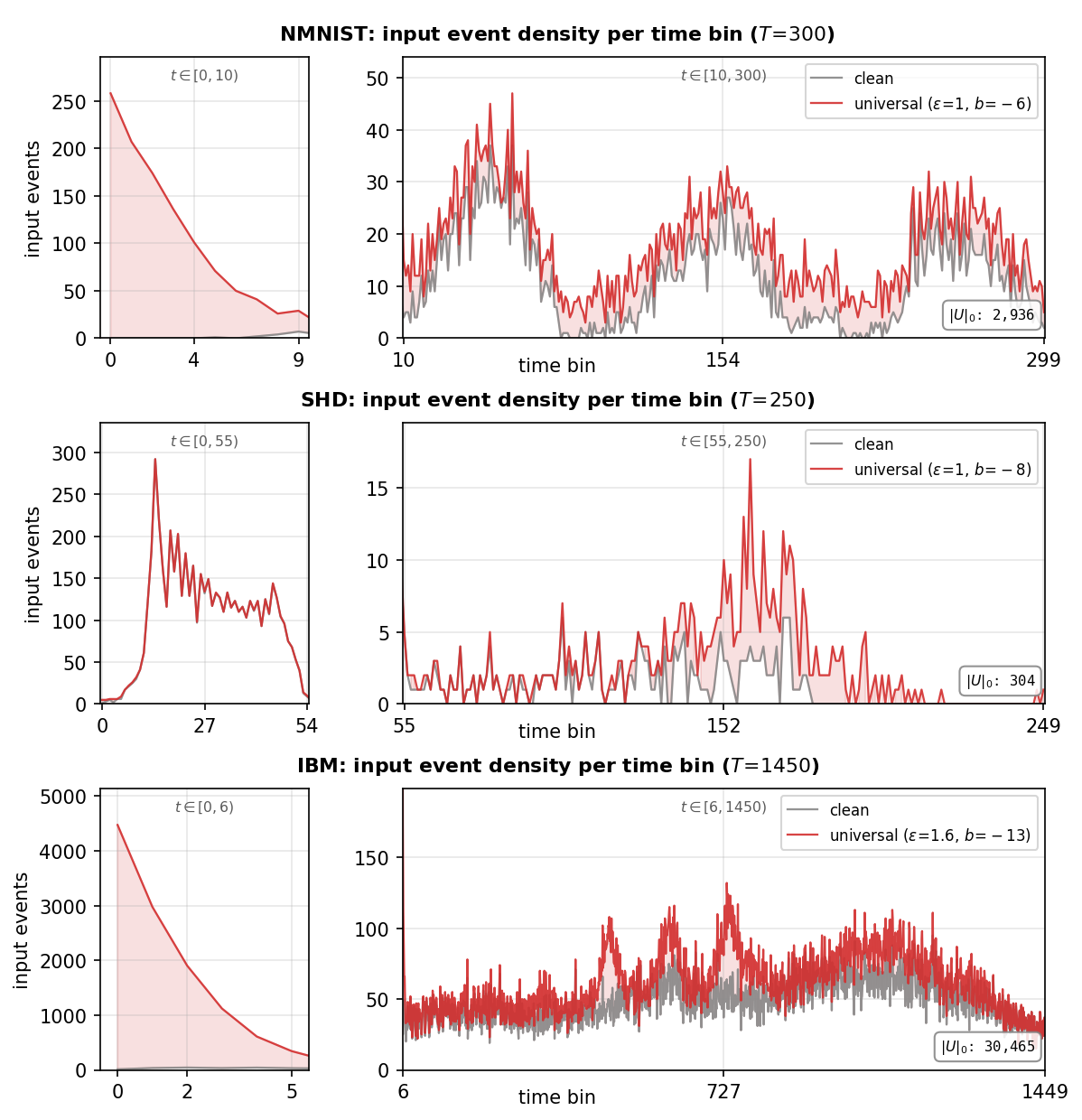}
\caption{Universal sponge: input-event count per time bin, clean
vs.\ sponged ($\mathbf{x}_0\,\oplus\,\mathbf{U}$).}
\label{fig:splitaxis_universal}
\end{figure}

\begin{figure}[t]
\centering
\includegraphics[width=\columnwidth]{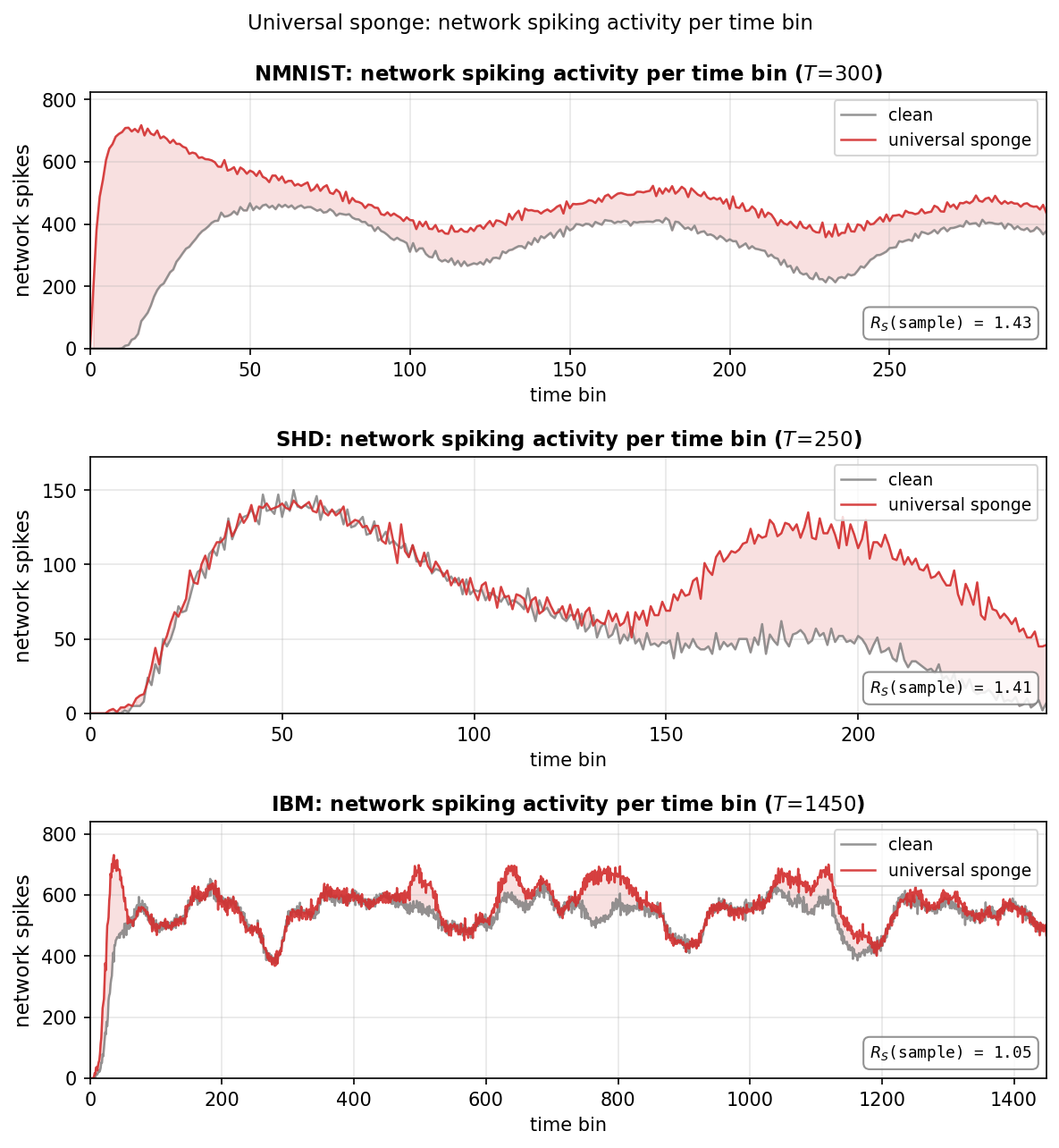}
\caption{Universal sponge: total network spikes per
time bin, clean vs.\ sponged ($\mathbf{x}_0\,\oplus\,\mathbf{U}$),
for a representative sample on each dataset.}
\label{fig:netact_universal}
\end{figure}

\begin{figure}[t]
\centering
\includegraphics[width=\columnwidth]{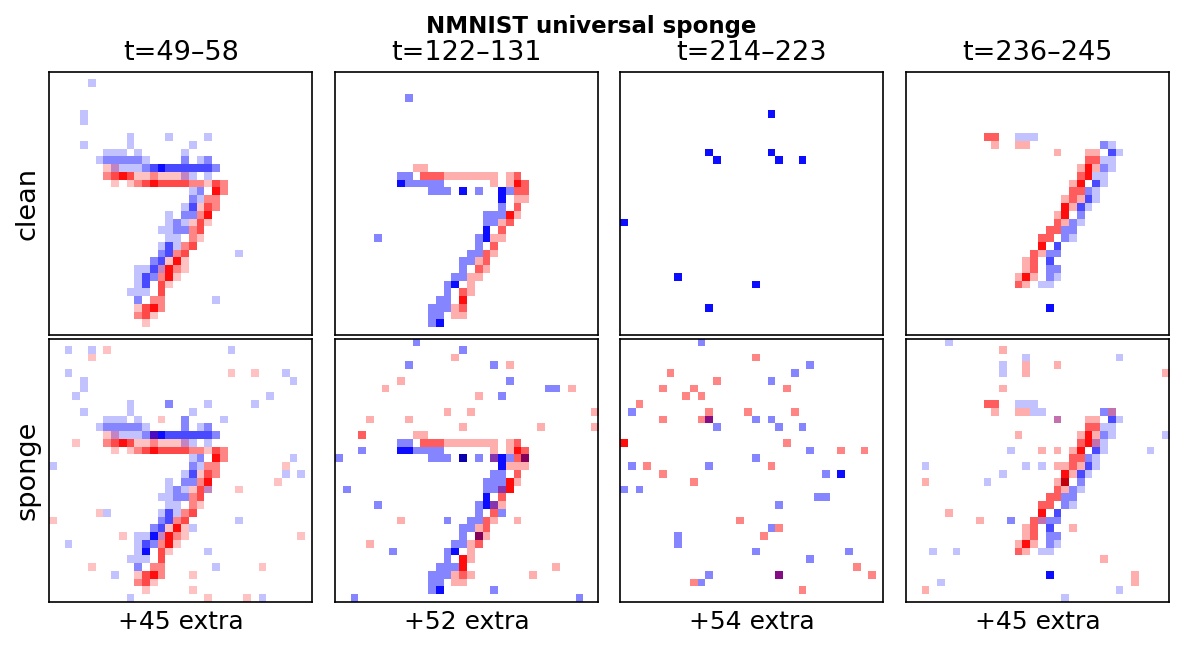}
\caption{NMNIST universal sponge: clean (top) vs.\ sponged
(bottom), four $10$-bin windows.}
\label{fig:lowpert_nmnist_universal}
\end{figure}

\begin{figure}[t]
\centering
\includegraphics[width=\columnwidth]{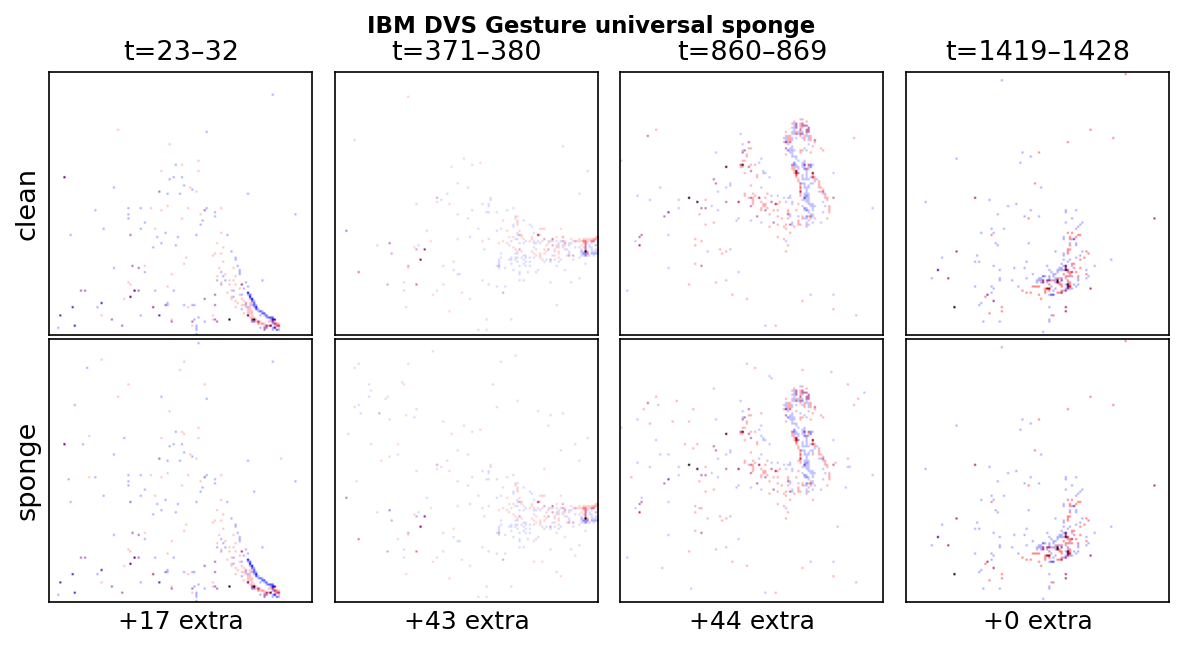}
\caption{IBM DVS Gesture universal sponge.}
\label{fig:lowpert_ibm_universal}
\end{figure}

\begin{figure}[t]
\centering
\begin{subfigure}{0.49\columnwidth}
  \includegraphics[width=\linewidth]{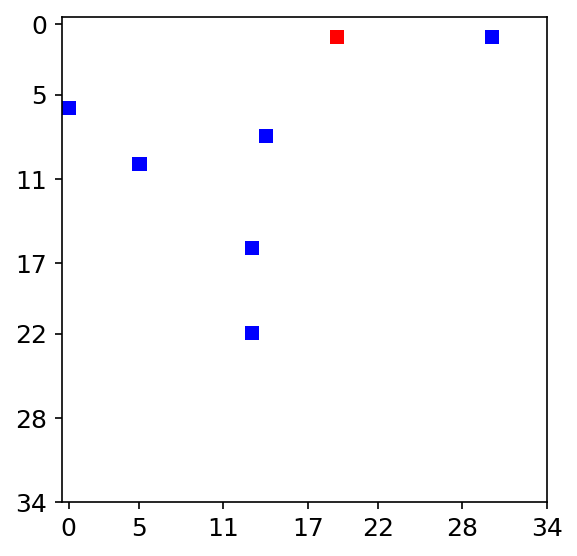}
  \caption{NMNIST ($t\!=\!213$)}\label{fig:ufilt_nmnist}
\end{subfigure}\hfill
\begin{subfigure}{0.49\columnwidth}
  \includegraphics[width=\linewidth]{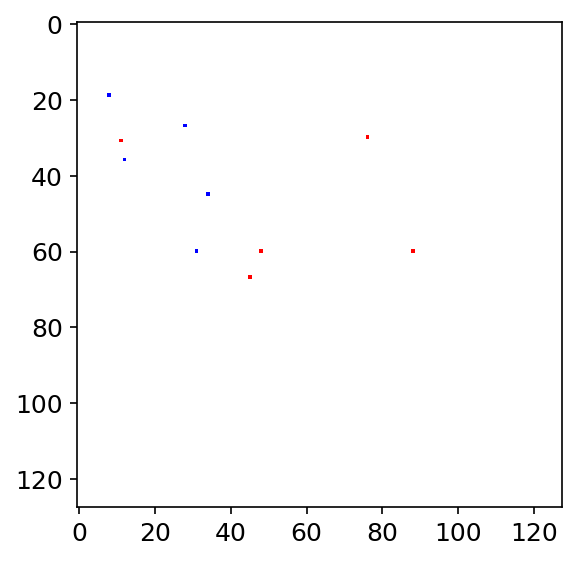}
  \caption{IBM ($t\!=\!131$)}\label{fig:ufilt_ibm}
\end{subfigure}

\vspace{0.4em}
\begin{subfigure}{\columnwidth}
  \includegraphics[width=\linewidth]{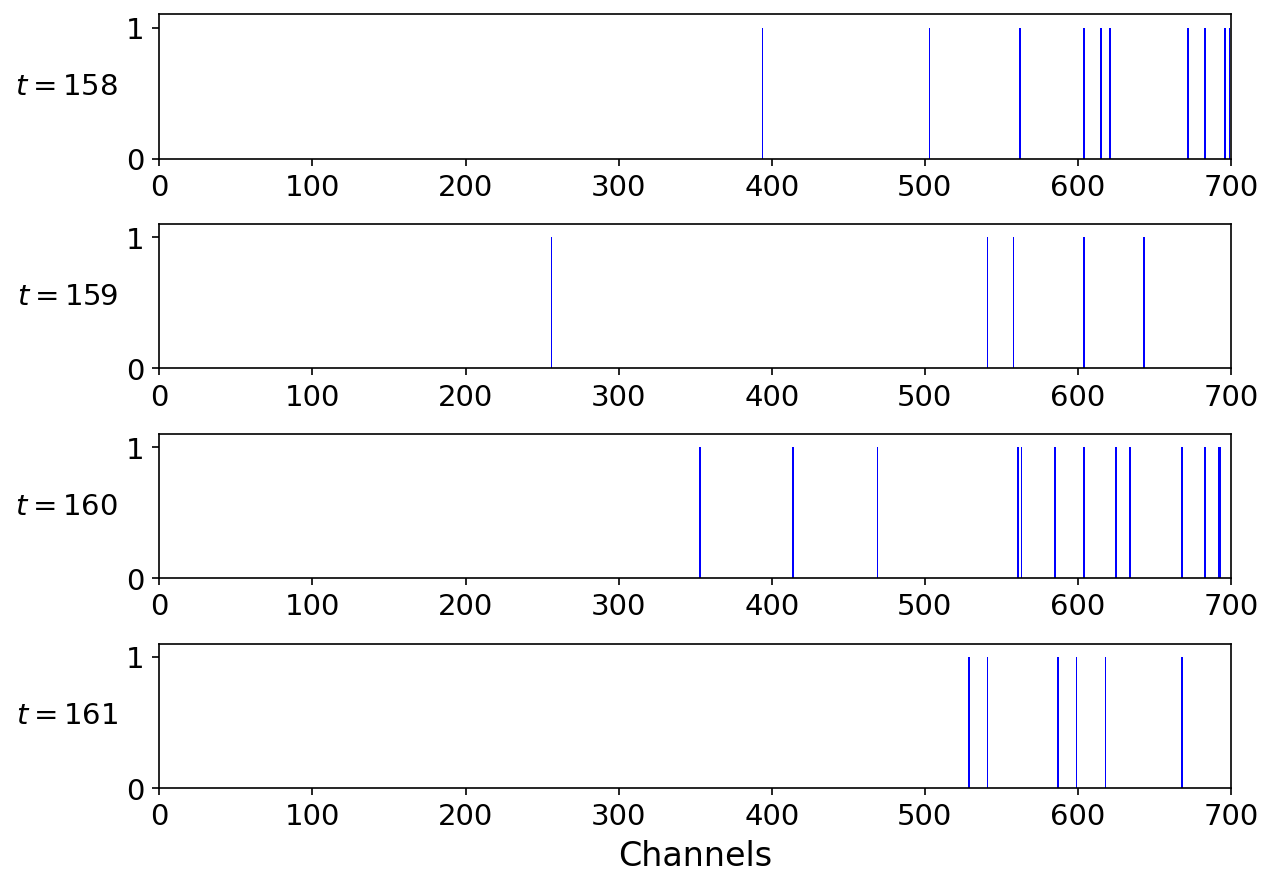}
  \caption{SHD (four consecutive bins)}\label{fig:ufilt_shd}
\end{subfigure}
\caption{Universal sponge masks $\mathbf{U}$ ($|\mathbf{U}|_0\!=\!2\,936$/$30\,465$/$304$ for
NMNIST/IBM/SHD), at one representative time bin (NMNIST/IBM,
blue\,=\,OFF, red\,=\,ON) or four consecutive bins (SHD).}
\label{fig:universal_filters}
\end{figure}

Figure~\ref{fig:splitaxis_universal} shows the universal sponge's
event distribution along $T$ (clean vs.\ clean\,$\oplus\,\mathbf{U}$),
with the same split time axis as
Figure~\ref{fig:splitaxis_persample}. Figure~\ref{fig:netact_universal} shows the network's internal
spiking activity per time bin under the clean and the sponged input
($\mathbf{x}_0\,\oplus\,\mathbf{U}$); the fixed mask drives a
sustained activity increase across the window, smaller than the
per-sample attack but at zero streaming-time cost. Figures~\ref{fig:lowpert_nmnist_universal}
and \ref{fig:lowpert_ibm_universal} show the sponged input for a
representative NMNIST and IBM sample.
Because $\mathbf{U}$ is a
fixed, dataset-independent tensor it can be visualized directly;
Figure~\ref{fig:universal_filters} shows a snapshot of the 
$\mathbf{U}$ for all three datasets.

Unlike the per-sample attack, a single fixed mask cannot track each input's decision boundary, so preservation is lower ($\Pi = 0.73$–$0.91$ — the clean prediction is kept on roughly $90\%$ of inputs for NMNIST and IBM) and sponge accuracy drops by $5.7$–$10.35$ pp. This is a deliberate trade: in exchange, the attack needs no per-input optimization, is injected once at the sensor, and raises per-inference energy across the whole input stream — independently of whether any individual prediction is preserved. Its advantage is operational reach and reusability at a bounded stealth cost, not the near-perfect preservation of the per-sample attack.

\subsection{Per-inference energy on Loihi 1}\label{sec:results-energy}
We translate the measured SynOp counts to estimated per-inference
energy on Intel's Loihi-1 chip~\cite{Loihi18} by multiplying
the SynOp count by $23.6\,\mathrm{pJ}$ per SynOp. Table~\ref{tab:loihi_energy} reports the absolute clean and sponge energies, and the per-inference $\Delta E$, for both per-sample and universal attack regimes.

\begin{table*}[t]
\centering
\caption{Per-inference Loihi-1 energy
$(\mathcal{E}_h\!=\!SynOps\!\cdot\! 23.6\,\mathrm{pJ})$ for
both attack regimes, all three datasets. Absolute energies should
be read with the first-order caveat that we ignore membrane-update
and static power. $\mathcal{E}_{\mathrm{clean}}$ and $\mathcal{E}_{\mathrm{sponge}}$ are per-inference means over the full test set. Dividing $\mathcal{E}_{\mathrm{sponge}}$ by $\mathcal{E}_{\mathrm{clean}}$ therefore yields a ratio of means, which differs slightly from the mean-of-per-sample-ratios $R_E$ reported in Table~\ref{tab:per_sample_main} when per-sample SynOps counts vary.}
\label{tab:loihi_energy}
\begin{tabular}{llrrr}
\toprule
SNN model & Regime & $\mathcal{E}_{\mathrm{clean}}$ & $\mathcal{E}_{\mathrm{sponge}}$ & $\Delta E$ \\
\midrule
\multirow{2}{*}{NMNIST}
        & per-sample        & $3.40\,\mathrm{mJ}$ & $5.06\,\mathrm{mJ}$ & $+1.66\,\mathrm{mJ}$ \\
        & universal sponge  & $3.40\,\mathrm{mJ}$ & $4.04\,\mathrm{mJ}$ & $+644\,\upmu\mathrm{J}$ \\
\midrule
\multirow{2}{*}{SHD}
        & per-sample        & $59\,\upmu\mathrm{J}$  & $93\,\upmu\mathrm{J}$ & $+34\,\upmu\mathrm{J}$ \\
        & universal sponge  & $59\,\upmu\mathrm{J}$  & $73\,\upmu\mathrm{J}$ & $+14\,\upmu\mathrm{J}$ \\
\midrule
\multirow{2}{*}{IBM DVS Gesture}
        & per-sample        & $8.98\,\mathrm{mJ}$ & $22.22\,\mathrm{mJ}$ & $+13.24\,\mathrm{mJ}$ \\
        & universal sponge  & $8.98\,\mathrm{mJ}$ & $9.79\,\mathrm{mJ}$  & $+808\,\upmu\mathrm{J}$ \\
\bottomrule
\end{tabular}
\end{table*}

\begin{figure}[t]
\centering
\includegraphics[width=\columnwidth]{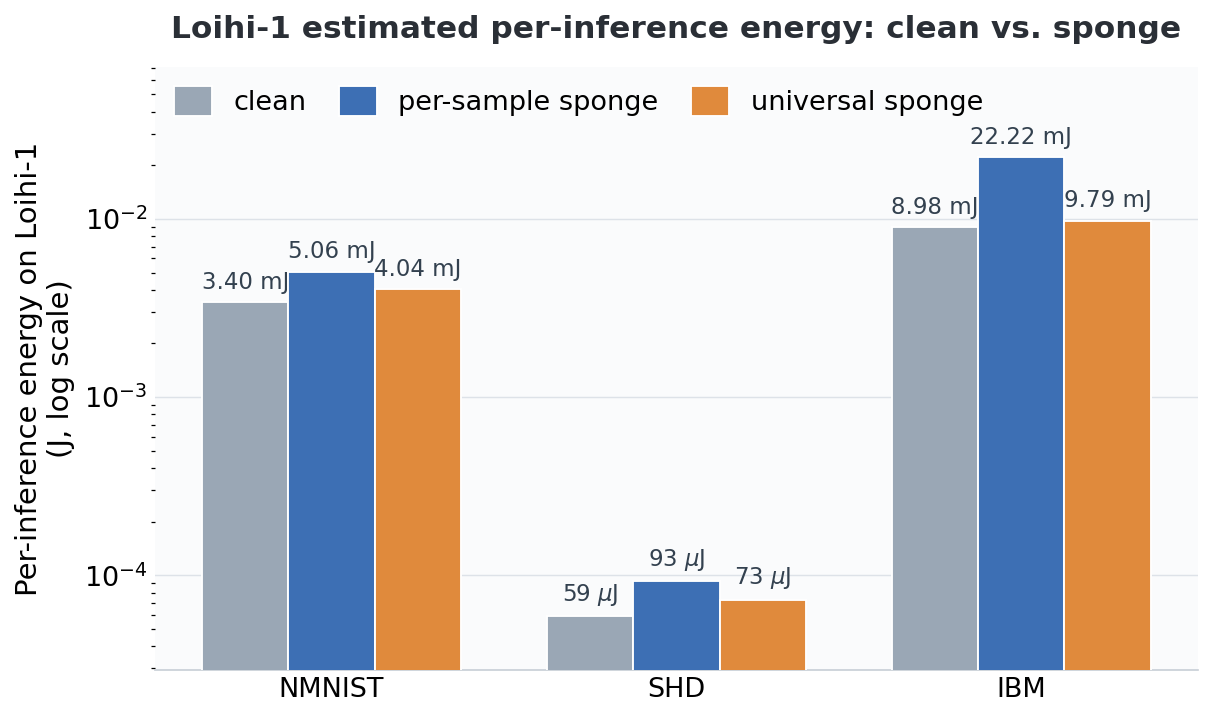}
\caption{Estimated per-inference energy on Loihi-1 (mean over the
full test set, log scale). Clean, per-sample, and universal sponge
bars per dataset; absolute values in
Table~\ref{tab:loihi_energy}.}
\label{fig:energy_bars}
\end{figure}

The absolute $\Delta E$ scales with input dimensionality even when
the inflation ratio is modest: IBM's universal-sponge has the smallest relative inflation of the three datasets, yet the
absolute $\Delta E$ ($+808\,\upmu\mathrm{J}$ per inference) is
larger in absolute terms than NMNIST's universal-sponge
$\Delta E$ ($+644\,\upmu\mathrm{J}$) and an order of magnitude
larger than SHD's. For a continuously-deployed device performing
one IBM-Gesture inference per second, the universal sponge
accumulates $\sim\!70\,\mathrm{J}$ per day of attack-driven battery
drain on Loihi-1; over a year of $24/7$ operation,
$\sim\!25.5\,\mathrm{kJ}$ ($\sim\!7\,\mathrm{Wh}$). Figure~\ref{fig:energy_bars} visualizes the estimated per-inference
energies on Loihi-1 across both attack regimes.

\subsection{Attack runtime}\label{sec:res-runtime}

\subsection*{Streaming feasibility of the per-sample attack}
The most operationally important property of the per-sample attack
is its end-to-end runtime per input. Table~\ref{tab:runtime}
reports the time at which the lowest-loss sponge input was empirically
found during the optimization running on an Nvidia A100. This is the
\emph{useful} attack time: the optimization could in principle stop
there with no loss in attack quality.

\begin{table}[t]
\centering
\caption{Per-sample attack runtime on an Nvidia A100. ``time-to-best''
is the wall time at which the lowest-loss
$\mathbf{x}_{\mathrm{sponge}}$ was found inside the optimization
loop.}
\label{tab:runtime}
\begin{tabular}{lc}
\toprule
SNN model & Time-to-best (mean) \\
\midrule
NMNIST          & $4.6\,\mathrm{s}$ \\
SHD             & $1.1\,\mathrm{s}$  \\
IBM DVS Gesture & $240.8\,\mathrm{s}$ ($\sim\!4\,\mathrm{min}$) \\
\bottomrule
\end{tabular}
\end{table}

Taking Table~\ref{tab:runtime} as the practical attack cost, the
per-sample attack on IBM costs $\sim\!4$ minutes per inference.
Compared to the inference rate a deployed device requires
(typically tens to hundreds of Hz), such a per-input optimization cost is difficult to reconcile with a streaming inference pipeline. This measurement makes concrete the ``baseline upper bound'' framing from Section~\ref{sec:threat}: the per-sample regime characterizes what input-space sponging can achieve, while the universal sponge is the operationally feasible realization.

\subsection*{Universal sponge runtime}
Constructing $\mathbf{U}$ is a one-shot pass over the $100$-$200$
training samples; on the A100 this takes about a minute on each
dataset. Per-inference application is an elementwise XOR with a
fixed binary mask: at most a few microseconds per inference on any
hardware that supports the bitwise operation, and effectively zero
attack-side compute on the device itself. Once $\mathbf{U}$ is
deployed (e.g.\ as a stuck-pixel pattern, a fixed light source in
the camera's field of view, or a compromised sensor driver), no
further attacker presence is required.

This is the key operational contrast: the per-sample attack
\emph{works in principle} but is difficult to hide behind the sensor of
a streaming deployment; the universal sponge is compatible with streaming deployment, and
delivers $43$--$74\%$ of the per-sample's spike inflation at zero
streaming-time attack cost.
\section{Conclusion} \label{sec:conclusion}

We studied input-space sponge attacks on Spiking Neural Networks in
their native binary event-based regime, the setting in which SNNs
are operationally valuable as low-energy edge inference engines.
We covered both the visual (DVS event camera) and the auditory
(silicon cochlea) sensing modalities, under two attacker
capabilities. The \emph{per-sample} attack inflates per-inference
SynOps by $1.5$ to $2.6\times$ across NMNIST, SHD, and IBM DVS
Gesture while preserving the predicted class on $\geq\!98\%$ of
each full test set, confirming that input-space sponge attacks
transfer from the rate-coded image regime to the binary event
regime that DVS cameras and silicon cochleae actually produce. Its
$\sim\!4$-minute per-inference optimization cost on IBM places it
as the achievable \emph{ceiling} of input-space sponging rather
than a streaming-time threat. The \emph{universal sponge}---to our
knowledge the first reported for SNNs in this binary event-based
regime---makes the sponge attack streaming-deployable at a bounded cost to inflation: it inflates SynOps by $1.09$ to
$1.24\times$ with one-time offline construction and constant-time
XOR application, and is realizable as a sensor-side overlay (a
fixed light pattern, a stationary acoustic signal, or a
compromised sensor driver) with no attacker presence required
once deployed.

Translating the measured SynOp counts to estimated Loihi-1 energy
yields per-inference deltas spanning $14\,\upmu\mathrm{J}$
(universal sponge, SHD) to $13.24\,\mathrm{mJ}$ (per-sample, IBM);
a continuously-deployed IBM-Gesture-class SNN under the universal
sponge alone would dissipate $\sim\!25.5\,\mathrm{kJ}$ per year of
attack-driven battery drain. Future work includes direct on-chip
energy measurement on Loihi and dedicated
defenses such as input-distribution monitoring and spike-rate
regularization. Of the two regimes, the per-sample attack
establishes the achievable ceiling, while the universal
sponge---deployable, reusable, and effectively free at
inference---is the one the neuromorphic community must plan against.


\bibliographystyle{IEEE}
\bibliography{refs_AI.bib, refs_AI_security, refs_SNN_security,}

\begin{thebibliography}{10}

\bibitem{Loihi18}
M.~Davies \textit{et al.},
\newblock ``Loihi: A neuromorphic manycore processor with on-chip learning,''
\newblock {\em {IEEE} Micro}, vol. 38, no. 1, pp. 82--99, Jan./Feb. 2018.

\bibitem{ASCA-IAMINDNTBBKMRJ15}
F.~Akopyan \textit{et al.},
\newblock ``True{N}orth: Design and tool flow of a 65 m{W} 1 million neuron programmable neurosynaptic chip,''
\newblock {\em IEEE Trans. Comput.-Aided Design Integr. Circuits Syst.}, vol. 34, no. 10, pp. 1537--1557, Oct. 2015.

\bibitem{shumailov2021sponge}
I.~Shumailov, Y.~Zhao, D.~Bates, N.~Papernot, R.~Mullins, and R.~Anderson,
\newblock ``Sponge examples: Energy-latency attacks on neural networks,''
\newblock in {\em Proc. IEEE Eur. Symp. Security and Privacy (EuroS\&P)}, 2021, pp. 212--231.

\bibitem{cina2022sponge}
A.~E. Cin{\`a}, A.~Demontis, B.~Biggio, F.~Roli, and M.~Pelillo,
\newblock ``Energy-latency attacks via sponge poisoning,''
\newblock {\em Information Sciences}, vol. 702, pp. 121905, Jun 2025.

\bibitem{krithivasan2022spikeattack}
S.~Krithivasan, S.~Sen, N.~Rathi, K.~Roy, and A.~Raghunathan,
\newblock ``Efficiency attacks on spiking neural networks,''
\newblock in {\em Proc. of the 59th ACM/IEEE Design Automation Conference (DAC)}, New York, USA, 2022, p. 373–378.

\bibitem{OJCT15}
G.~Orchard, A.~Jayawant, G.~K. Cohen, and N.~Thakor,
\newblock ``Converting static image datasets to spiking neuromorphic datasets using saccades,''
\newblock {\em Front. Neurosci.}, vol. 9, Nov. 2015,
\newblock Article 437.

\bibitem{heidelberg}
B.~Cramer, Y.~Stradmann, J.~Schemmel, and F.~Zenke,
\newblock ``The {H}eidelberg spiking data sets for the systematic evaluation of spiking neural networks,''
\newblock {\em IEEE Trans. Neural Netw. Learn. Syst.}, vol. 33, no. 7, pp. 2744--2757, Jul. 2022.

\bibitem{ATBM17}
A.~Amir \textit{et al.},
\newblock ``A low power, fully event-based gesture recognition system,''
\newblock in {\em Proc. IEEE Conf. Comput. Vis. Pattern Recognit. (CVPR)}, Jul. 2017.

\bibitem{shapira2023phantom}
A.~Shapira, A.~Zolfi, L.~Demetrio, B.~Biggio, and A.~Shabtai,
\newblock ``Phantom sponges: Exploiting non-maximum suppression to attack deep object detectors,''
\newblock in {\em Proc. IEEE/CVF Winter Conf. on Applications of Computer Vision (WACV)}, Jan 2023, pp. 4560--4569.

\bibitem{hong2021earlyexit}
S.~Hong, Y.~Kaya, I.-V. Modoranu, and T.~Dumitra{\c{s}},
\newblock ``A {P}anda? {N}o, {I}t's a {S}loth: {S}lowdown {A}ttacks on {A}daptive {M}ulti-{E}xit {N}eural {N}etwork {I}nference,''
\newblock in {\em Int. Conf. on Learning Representations (ICLR)}, 2021.

\bibitem{telintelo2024skipsponge}
J.~te~Lintelo, S.~Koffas, and S.~Picek,
\newblock ``The {SkipSponge} attack: Sponge weight poisoning of deep neural networks,''
\newblock {\em ITU Journal on Future and Evolving Technologies}, vol. 6, no. 3, Sept. 2025.

\bibitem{11295756}
N.~Yang et~al.,
\newblock ``{EOS}: An energy-oriented attack framework for spiking neural networks,''
\newblock in {\em 61st ACM/IEEE Design Automation Conference (DAC)}, 2024, pp. 1--6.

\bibitem{BaSiRa18}
A.~Bagheri, O.~Simeone, and B.~Rajendran,
\newblock ``Adversarial training for probabilistic spiking neural networks,''
\newblock in {\em Proc. IEEE Int. Workshop Signal Process. Adv. Wireless Commun. (SPAWC)}, Jun. 2018.

\bibitem{BLHSS18}
J.~Büchel, G.~Lenz, Y.~Hu, S.~Sheik, and M.~Sorbaro,
\newblock ``Adversarial attacks on spiking convolutional neural networks for event-based vision,''
\newblock {\em Front. Neurosci.}, vol. 16, Dec. 2022.

\bibitem{SPSLPR19}
S.~Sharmin, P.~Panda, S.~S. Sarwar, C.~Lee, W.~Ponghiran, and K.~Roy,
\newblock ``A comprehensive analysis on adversarial robustness of spiking neural networks,''
\newblock in {\em Proc. Int. Jt. Conf. Neural Netw. (IJCNN)}, Jul. 2019.

\bibitem{E-AMSA21}
R.~El-Allami, A.~Marchisio, M.~Shafique, and I.~Alouani,
\newblock ``Securing deep spiking neural networks against adversarial attacks through inherent structural parameters,''
\newblock in {\em Proc. Design Autom. Test Europe Conf. (DATE)}, Feb. 2021, pp. 774--779.

\bibitem{MPMMS21}
A.~Marchisio, G.~Pira, M.~Martina, G.~Masera, and M.~Shafique,
\newblock ``{DVS}-{A}ttacks: Adversarial attacks on dynamic vision sensors for spiking neural networks,''
\newblock in {\em Proc. Int. Jt. Conf. Neural Netw. (IJCNN)}, Jul. 2021.

\bibitem{LHDWLDLX23}
L.~Liang \textit{et al.},
\newblock ``Exploring adversarial attack in spiking neural networks with spike-compatible gradient,''
\newblock {\em IEEE Trans. Neural Netw. Learn. Syst.}, vol. 34, no. 5, pp. 2569--2583, May 2023.

\bibitem{MNKHMS20}
A.~Marchisio, G.~Nanfa, F.~Khalid, M.~A. Hanif, M.~Martina, and M.~Shafique,
\newblock ``Is spiking secure? a comparative study on the security vulnerabilities of spiking and deep neural networks,''
\newblock in {\em Proc. Int. Jt. Conf. Neural Netw. (IJCNN)}, Jul. 2020.

\bibitem{NSHM22}
O.~Nomura, Y.~Sakemi, T.~Hosomi, and T.~Morie,
\newblock ``Robustness of spiking neural networks based on time-to-first-spike encoding against adversarial attacks,''
\newblock {\em IEEE Trans. Circuits Syst. II: Express Br.}, vol. 69, no. 9, pp. 3640--3644, Sep. 2022.

\bibitem{RaSt25b}
S.~Raptis and H.~G. Stratigopoulos,
\newblock ``Input-specific and universal adversarial attack generation for spiking neural networks in the spiking domain,''
\newblock in {\em Proc. Int. Jt. Conf. Neural Netw. (IJCNN)}, Jun./Jul. 2025.

\bibitem{RaSt26c}
S.~Raptis and H.-G. Stratigopoulos,
\newblock ``Adversarial vulnerability of neuromorphic processors demonstrated on loihi chip,''
\newblock in {\em Proc. IEEE Int. Conf. Artif. Intell. Circuits Syst. (AICAS)}, Sep. 2026.

\bibitem{VMAMS20}
V.~Venceslai, A.~Marchisio, I.~Alouani, M.~Martina, and M.~Shafique,
\newblock ``Neuroattack: Undermining spiking neural networks security through externally triggered bit-flips,''
\newblock in {\em Proc. Int. Jt. Conf. Neural Netw. (IJCNN)}, Jul. 2020.

\bibitem{RKKAS25}
S.~Raptis, P.~Kling, I.~Kaskampas, I.~Alouani, and H.-G. Stratigopoulos,
\newblock ``Input-triggered hardware trojan attack on spiking neural networks,''
\newblock in {\em Proc. IEEE Int. Symp. Hardw.-Oriented Secur. Trust (HOST)}, May 2025.

\bibitem{BD-RADARS26}
V.~Barbaza et~al.,
\newblock ``Stealing {AI} model weights through covert communication channels,''
\newblock {\em IEEE Trans. Very Large Scale Integr. ({VLSI}) Syst.}, 2026.

\bibitem{RaSt26d}
S.~Raptis and H.-G. Stratigopoulos,
\newblock ``Training-free intellectual property protection for spiking neural networks,''
\newblock in {\em Proc. Int. Jt. Conf. Neural Netw. (IJCNN)}, Jun. 2026.

\bibitem{AEPU24}
G.~Abad, O.~Ersoy, S.~Picek, and A.~Urbieta,
\newblock ``Sneaky spikes: Uncovering stealthy backdoor attacks in spiking neural networks with neuromorphic data,''
\newblock in {\em Proc. Symp. Netw. Distrib. Syst. Secur. (NDSS)}, Feb. 2024.

\bibitem{10.1007/978-981-96-7005-5_17}
H.~Fu, G.~Li, J.~Wu, J.~Li, K.~Zhou, and Y.~Liu,
\newblock ``Spikewhisper: Temporal spike backdoor attacks on federated neuromorphic learning over low-power devices,''
\newblock in {\em Neural Information Processing}, M.~Mahmud, M.~Doborjeh, K.~Wong, A.~C.~S. Leung, Z.~Doborjeh, and M.~Tanveer, Eds., Singapore, 2025, pp. 243--258, Springer Nature Singapore.

\bibitem{10.1007/978-3-032-07884-1_1}
G.~Abad, S.~Picek, and A.~Urbieta,
\newblock ``Time-distributed backdoor attacks on federated spiking learning,''
\newblock in {\em Computer Security -- ESORICS 2025}, V.~Nicomette, A.~Benzekri, N.~Boulahia-Cuppens, and J.~Vaidya, Eds., Cham, 2026, pp. 1--20, Springer Nature Switzerland.

\bibitem{11391974}
R.~Riaño, G.~Abad, S.~Picek, and A.~Urbieta,
\newblock ``Flashy backdoor:proc. annu. comput. secur. appl. conf. real-world environment backdoor attack on {SNN}s with {DVS} cameras,''
\newblock in {\em IEEE Proc. Annu. Comput. Secur. Appl. Conf. (ACSAC)}, 2025, pp. 986--1002.

\bibitem{NLEKKG22}
K.~Nagarajan, J.~Li, S.~S. Ensan, M.~N.~I. Khan, S.~Kannan, and S.~Ghosh,
\newblock ``Analysis of power-oriented fault injection attacks on spiking neural networks,''
\newblock in {\em Proc. Design Autom. Test Europe Conf. (DATE)}, Mar. 2022, pp. 861--866.

\bibitem{NRTKG23}
K.~Nagarajan, R.~Roy, R.~O. Topaloglu, S.~Kannan, and S.~Ghosh,
\newblock ``{SCANN}: Side channel analysis of spiking neural networks,''
\newblock {\em Cryptography}, vol. 7, no. 2, Mar. 2023.

\bibitem{GoDaSu24}
B.~Goswami, T.~Das, and M.~Suri,
\newblock ``Experimental investigation of side-channel attacks on neuromorphic spiking neural networks,''
\newblock {\em IEEE Embed. Syst. Lett.}, vol. 16, no. 2, pp. 231--234, Jun. 2024.

\bibitem{aksu2025privacyfederatedlearningspiking}
D.~Aksu, J.~M. del Rincon, and I.~Alouani,
\newblock ``Privacy in federated learning with spiking neural networks,''
\newblock {\em arXiv:2511.21181}, 2025.

\bibitem{Adam2017optimization}
D.~P. Kingma and J.~Ba,
\newblock ``Adam: A method for stochastic optimization,''
\newblock {\em arXiv:1412.6980}, 2017.

\bibitem{shor18}
S.~B. Shrestha and G.~Orchard,
\newblock ``{SLAYER}: Spike layer error reassignment in time,''
\newblock in {\em Proc. Adv. Neural Inf. Process. Syst. (NeurIPS)}, Dec. 2018, pp. 1412--1421.

\end{thebibliography}

\end{document}